\documentclass[twoside,amsmath,amssymb,aps,pre,reprint]{revtex4-2}
\usepackage[utf8]{inputenc}
\usepackage[T1]{fontenc}
\usepackage{newtxtext,newtxmath}

\usepackage{xcolor,graphicx}
\definecolor{TiffanyBlue}{cmyk}{1, 0, 0.3, 0}

\usepackage[
    pdftitle={Article},
    pdfauthor={Author},
    bookmarks=true,
    colorlinks=true,
    linkcolor=blue,
    urlcolor=blue,
    citecolor=TiffanyBlue,
    plainpages=false,
    pdfpagelabels,
    final,
    breaklinks=true
]{hyperref}
\usepackage{orcidlink}
\usepackage{fancyhdr}

\usepackage{breakurl}
\usepackage{mathtools,bm,etoolbox,empheq}
\usepackage{enumitem,placeins}
\usepackage{booktabs,bookmark}
\usepackage{lipsum}

\DeclareRobustCommand{\bmk}[1]{\bm{#1}}
\setlist[itemize,enumerate,description]{leftmargin=*}

\newcommand{\ko}[1]{\left( #1 \right)}
\newcommand{\kko}[1]{\left[ #1 \right]}

\newcommand{\abs}[1]{\left| #1 \right|}
\newcommand{\dd}{\mathop{}\!d}
\newcommand{\dD}{\mathop{}\!\Delta}
\newcommand{\Expt}{\mathbb{E}}
\newcommand{\Var}{\operatorname{Var}}
\newcommand{\SD}{\operatorname{SD}}

\DeclarePairedDelimiter\ceil{\lceil}{\rceil}

\def\no{\nonumber}

\fancypagestyle{firstpage}{
  \fancyhf{}
  \fancyhead[C]{\large PHYSICAL REVIEW E \textbf{112}, 044304 (2025)}
  \fancyhead[RE,LO,LE,RO]{}
  \fancyfoot[C]{\small 044304-\thepage}
  \fancyfoot[RE,LO,LE,RO]{}}

\fancypagestyle{restpage}{
  \fancyhf{}
  \fancyhead[LO]{\small FRACTIONAL STOCHASTIC MODEL OF CITATION DYNAMICS \ldots}
  \fancyhead[LE]{\small KEISUKE OKAMURA}
  \fancyhead[RE]{\small PHYSICAL REVIEW E \textbf{112}, 044304 (2025)}
  \fancyhead[RO]{\small PHYSICAL REVIEW E \textbf{112}, 044304 (2025)}
  \fancyfoot[C]{\small 044304-\thepage}}

\fancypagestyle{suppl1}{
  \fancyhf{}
  \fancyhead[C]{\large PHYSICAL REVIEW E \textbf{112}, 044304 (2025)}
  \fancyfoot[C]{\small 044304-suppl-\thepage}
  }

\fancypagestyle{suppl2}{
  \fancyhf{}
  \fancyfoot[C]{\small 044304-suppl-\thepage}
  }

\begin{document}

\def\mtitle{%
%
Fractional stochastic model of citation dynamics with memory and volatility}
%
\title{\vspace*{0.4em}\mtitle}

\author{Keisuke Okamura\,\orcidlink{0000-0002-0988-6392}}
\email[Contact author:~]{okamura@alumni.lse.ac.uk}
\affiliation{Embassy of Japan in the United States of America, Washington, DC, USA}
\affiliation{SciREX Center, National Graduate Institute for Policy Studies, Tokyo, Japan}

\makeatletter
\def\@date{Submitted 6 April 2025; accepted 29 August 2025; published 9 October 2025}
\makeatother

\begin{abstract}
Understanding the statistical laws governing citation dynamics remains a fundamental challenge in network theory and the science of science. Citation networks typically exhibit in-degree distributions well approximated by log-normal distributions yet also display power-law behaviour in the high-citation regime---an apparent contradiction lacking a unified explanation. Here we identify a previously unrecognised phenomenon: the variance of the logarithm of citation counts per unit time follows a power law with respect to time ($t$) since publication, scaling as $t^{H}$, with $H$ constant. This discovery introduces a new challenge while simultaneously offering a crucial clue to resolving this discrepancy. We develop a stochastic model in which latent attention to publications evolves through a memory-driven process with cumulative advantage, modelled as fractional Brownian motion with Hurst parameter $H$ and volatility. We show that antipersistent fluctuations in attention ($H < 1/2$) yield log-normal citation distributions, whereas persistent attention dynamics ($H > 1/2$) favour heavy-tailed power laws, thus resolving the log-normal--power-law contradiction. Numerical simulations confirm both the $t^{H}$ law and the transition between regimes. Empirical analysis of arXiv e-prints indicates that the latent attention process is intrinsically antipersistent ($H \approx 0.13$). By linking memory effects and stochastic fluctuations in attention to broader network dynamics, our findings provide a unifying framework for understanding the evolution of collective attention in science and other attention-driven processes.\\[1em]
DOI: \href {https://doi.org/10.1103/l2xd-43n9}{10.1103/l2xd-43n9}
\end{abstract}


\maketitle

\thispagestyle{firstpage}
\pagestyle{restpage}

\section{Introduction\label{sec:intro}}

The study of stochastic fluctuations in physics, originating from Einstein's seminal investigation of Brownian motion \cite{Einstein05}, has reshaped our understanding of the natural world.
Often unpredictable at the microscopic level, these dynamics have revealed fundamental parameters and governing principles through collective behaviour. 
This realisation has led to the recognition that many physical phenomena can be described within a universal framework, independent of their microscopic details. 
These insights have been pivotal in advancing nonequilibrium statistical mechanics---encompassing nonlinear dynamics and network theory---and have influenced chaos and fractal theory \cite{Mandelbrot82,Falconer14}.
Beyond physics and mathematics, they have helped explain biological morphogenesis \cite{Turing52,*Keller71}, the structural and dynamic evolution of large-scale networks such as the Internet, and complex social phenomena \cite{Watts98,*Albert02}.

Among the large-scale networks that exemplify these principles, a particularly relevant example arises in academia itself.  
In this domain, the scholarly practice of citation, where researchers systematically reference prior work to position their own findings, gives rise to a citation network \cite{Price65}.  
This network is constructed through the cumulative assembly of references and bibliographies across a vast collection of publications.  
Over the course of scientific history, this citation network has evolved into a substantial directed graph with a distinctive structure.  

Extensive research has examined how citation network structures emerge and evolve, offering insights into the self-organising principles of complex systems.  
Studies in this field have primarily focused on two key areas.  
The first concerns the distribution of cumulative citation counts, which corresponds to the in-degree distribution in network growth models.  
For decades, researchers have sought to determine the most suitable functional form to describe this distribution.  
Empirical studies have predominantly found that citation distributions are best fitted either by a scale-free (shifted) power law \cite{Price76,Redner98,Hajra06,Lehmann03,*Clauset09,*Peterson10,*Brzezinski15,Eom11} or by a log-normal distribution \cite{Redner05,Radicchi08,*Evans12,*Chatterjee16,*DAngelo19,Golosovsky17,Okamura22,Thelwall16b}.  
A key question remains whether a form of universality exists within these distributions \cite{Radicchi08,*Evans12,*Chatterjee16,*DAngelo19,Waltman12,*Golosovsky21}.

Another fundamental aspect of citation network research concerns its temporal evolution, specifically how the number of citations to individual publications changes over time \cite{Price65,Line74,*Nakamoto88,*Lariviere08,*Parolo15,*Yin17,Golosovsky12b,Golosovsky12,Wang13,Golosovsky17}.
Most publications receive few or no citations \cite{Price65,Seglen92,Redner98}.
Among those that do, some gain recognition only long after publication \cite{Redner05,Raan04,*Ke15}, while others continue accumulating citations indefinitely.
Each case follows a distinct citation pattern yet collectively they exhibit a characteristic `jump-and-decay' trend: citations rise sharply after publication, peak briefly, and then gradually decline, asymptotically approaching zero with a slower descent than the initial rise.

Between the two perspectives discussed above, the first, which examines the in-degree distribution, offers a `spatial' snapshot of citation network growth.
In contrast, the second, which analyses temporal citation patterns, traces the trajectory of a single point within this space or the evolution of statistical measures such as the mean citation count.  
A deeper understanding of the spatiotemporal properties of citation dynamics and their underlying drivers is not only essential in its own right but also provides valuable insights for analogous network growth models.

A paradigmatic mechanism often invoked to explain citation dynamics is cumulative advantage \cite{Price76}, which has played a foundational role in modelling efforts, despite its known limitations in reproducing detailed structures of real networks.
This principle refers to the idea that entities with an initial advantage---such as early citations---tend to attract further attention, thereby accumulating disproportionately more citations over time.
In networks, preferential attachment \cite{Barabasi99,*Barabasi02,*Jeong03,*Newman09} formalises this idea by favouring nodes with a greater number of links.
When applied linearly, it leads to a power-law distribution.
Furthermore, Gibrat's law \cite{Sutton97} extends linear preferential attachment to a continuous, stochastic setting.
Historically, these concepts have been framed as the Yule process \cite{Yule25,*Simon55} or, more commonly, the Matthew effect, often summarised as the `rich get richer' phenomenon.

A key question in citation network research is whether a theoretical framework can be developed that accommodates diverse temporal citation trajectories at the individual level while collectively reproducing the characteristic in-degree distribution and incorporating preferential attachment.
Addressing this question requires recognising a previously overlooked empirical observation: the standard deviation (SD) of the logarithm of citation counts evolves as a power law with respect to time since publication (the $t^H$ law; see Sec.~\ref{sec:preliminary}).
Existing theories of citation dynamics do not naturally account for this phenomenon.

To bridge this gap, we propose a theoretical model that unifies and explains a range of empirical findings within a coherent mathematical framework (Sec.~\ref{sec:theory}).
The validity of our theory is first examined through numerical simulations, which assess both its explanatory strength and predictive capability (Sec.~\ref{sec:simulation}).
We then test the model against empirical data to evaluate its effectiveness in describing real-world citation networks (Sec.~\ref{sec:analysis}).
Finally, we explore the broader implications of our framework (Sec.~\ref{sec:discussion}).

Our approach offers fresh insight into why in-degree distributions in citation networks are often well approximated by log-normal forms while simultaneously exhibiting power-law behaviour in the high-citation regime.
It also uncovers an intrinsic fractal structure underlying citation trajectories.
At the heart of our theory lies a fundamental principle introduced at the outset: the science of stochastic fluctuations.

\section{Preliminary Observations\label{sec:preliminary}}

In this section, we first provide a brief review of previous approaches to citation dynamics.
We then introduce a new empirical finding that remains unexplained by existing theories, underscoring the need for a more comprehensive theoretical framework.

\subsection{`\textit{Scientometric Engineering}'\label{sec:scienteng}}

Research on citation dynamics has primarily aimed to explain and predict the temporal trajectory of collective attention to science and its long-term impact.
Various approaches have been proposed to address this challenge.
One notable approach adopts an explanatory perspective, utilising a publication's citation history up to a given point, along with intrinsic characteristics such as fitness, to characterise its future citation trajectory \cite{Wang13}.
In contrast, a more predictive approach leverages bibliometric features, including a publication's formal attributes, novelty, originality and significance, alongside early citation counts from the first few years after publication, to forecast future citation behaviour \cite{Uzzi13,*Stegehuis15,*Cao16}.
Other stochastic approaches have also been proposed to enhance understanding of the citation process \cite{Glanzel95b,*Klemm02,Hajra06,Eom11,Golosovsky12b,Golosovsky12,Golosovsky17}.
While these modelling efforts achieve varying degrees of success depending on their objectives, a definitive formulation of citation dynamics has yet to be established.

Naturally, predicting the exact citation history of an individual publication remains impossible.
Even models incorporating pre-publication attributes or early citation patterns cannot perfectly replicate its time evolution.
The inherently creative, serendipitous, and collaborative nature of scientific progress introduces fundamental uncertainty, which any formulation of citation dynamics must explicitly address.
In this context, a stochastic approach indeed offers a natural framework for describing the citation process.
However, adopting a probabilistic perspective on a physical phenomenon requires careful examination of how randomness manifests, which stochastic process governs its dynamics, and what underlying mechanisms drive this uncertainty.

An insightful analogy is the rolling of a die \cite{Tasaki05}.
Before impact, its motion can be predicted and controlled to some extent, as it follows the deterministic movement of its centre of mass.
However, once it makes contact with a surface, the outcome becomes unpredictable, governed by complex rotational dynamics influenced by the specific conditions of impact.  
Thus, the result of a die roll comprises two components: one deterministic---predictable and controllable---and the other stochastic.
Together, these components shape the probability distribution of outcomes when the die is rolled repeatedly.

A similar principle governs collective attention to science, as reflected in citation patterns, where both a predictable, controllable component and an inherently random component coexist.
In this context, the `die' represents not an individual publication but a broader entity, such as science as a whole or a specific research community in disciplines like physics or mathematics, where potential attention is distributed.
The deterministic aspect of citation dynamics is shaped by quantifiable factors, including whether the author is an established scholar, the prominence of the publication venue (e.g., journal impact factor), prior citations and attention received, and intrinsic attributes such as fitness or novelty, provided they are properly quantified \cite{Wang13,Uzzi13,*Stegehuis15,*Cao16}.
In contrast, stochastic elements arise from interactions with the citing community---researchers who may reference the work.
These factors include cognitive limitations, resource constraints, time pressures, and the serendipitous discovery of new connections to other studies postpublication.
Such unpredictable influences contribute to the inherent stochasticity of the citation process.

Historically, certain fields have advanced by embracing this principle---integrating deterministic and stochastic elements into their theoretical frameworks through mathematical formalisation.
A prominent example is financial engineering, where the modelling of stochastic fluctuations has provided deep insights into the temporal evolution of risky asset prices in economics and finance.
A quintessential classical example is the Black--Scholes model \cite{Black73,*Merton73}, which utilises a stochastic differential equation (SDE) to describe asset price movements over time.
By modelling price dynamics as geometric Brownian motion, it establishes a theoretical foundation for option pricing.

More recently, a comparable strategy has been proposed for elucidating citation dynamics \cite{Okamura22}.
This work hypothesised that citation dynamics could similarly be modelled using an SDE, incorporating both deterministic and stochastic components, as in the Black--Scholes model.
It led to the formulation of an SDE for citation dynamics based on a generalised version of Black--Scholes' geometric Brownian motion.
Yet a crucial empirical phenomenon in citation dynamics, while hinted at in that study (Appendix~C, Supplemental Fig.~S9 in Ref.~\cite{Okamura22}), was not explicitly emphasised \cite{suppletext}.
As detailed below, this characteristic concerns the relationship between citation counts per unit time and the elapsed time since publication.

\subsection{The $\bmk{t^{H}}$ law\label{sec:law}}

\begin{figure}
\centering
\includegraphics[width=1.0\linewidth]{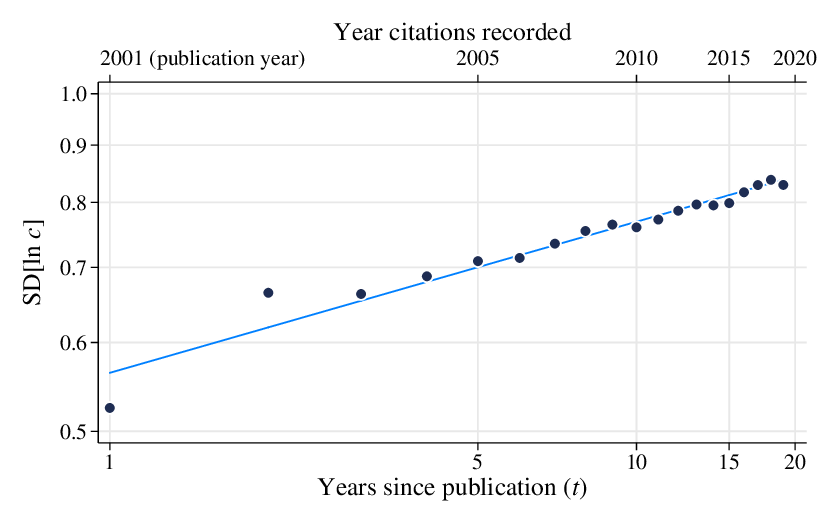}
\caption{The relationship between the standard deviation (SD) of the logarithm of annual citation counts ($c$) and the number of years since publication ($t$).
The data are based on Ref.~\cite{Okamura21}, focusing on e-prints published on arXiv in 2001 ($N=39{,}106$).}
\label{fig:t^H}
\end{figure}

Figure~\ref{fig:t^H} illustrates the relationship between the SD of the logarithm of annual citation counts ($c$) and the number of years since posting ($t$) for e-prints published on arXiv \cite{arXiv} in 2001 ($N=39{,}106$).
As is evident from the figure, SD increases with $t$ following a power-law relationship with a positive exponent $H$:
\begin{equation}\label{law:t^H}
\SD[\ln c]~\sim~t^{H}\,.
\end{equation}
This relationship holds consistently across publications from different publication years, as well as when the analysis is confined to specific research fields rather than the entire arXiv.
It also remains robust when alternative methods of conceptualising elapsed time are employed \cite{suppletext}.
Given this apparent universality, this paper refers to Eq.~\eqref{law:t^H} as the $t^H$ law.
For the dataset in Fig.~\ref{fig:t^H}, a linear regression analysis estimates the exponent as $\hat{H}\approx 0.13\pm 0.016$ $(p<0.001)$.
A more detailed analysis and discussion are presented later in Sec.~\ref{sec:analysis}; for now, the empirical findings above already provide strong motivation for developing a theoretical framework with enhanced explanatory power.

A previous study \cite{Okamura22}, using the same dataset as Fig.~\ref{fig:t^H}, concluded that the cumulative citation distribution for arXiv e-prints is better characterised by a log-normal distribution than a (shifted) power law.
Notably, the shape parameter ($\sigma$) of the log-normal distribution does not directly correspond to $\SD[\ln c]$ in Eq.~\eqref{law:t^H}.
This discrepancy arises because $\sigma$ characterises the distribution of cumulative citation counts over a publication's lifetime or a long-term period of interest, whereas $\SD[\ln c]$ reflects citation counts within a short, fixed time window (in this case, each year).
The two quantities are connected through a time-integrated relationship, both increasing monotonically with time since publication.
Indeed, an empirical relationship, $\sigma\sim t^{H'}$, with $H'$ constant, has been observed \cite{Okamura22}.
This trend can be interpreted as an effective law arising from the `convolution' of the $t^{H}$ law in Eq.~\eqref{law:t^H}.

In previous studies, when the cumulative distribution was identified as log-normal, the parameter $\sigma$ was typically treated as a static quantity. 
Although its estimated value varies across datasets and research fields, it is generally reported to range from 1 to 1.4 \cite{Radicchi08,*Evans12,*Chatterjee16,*DAngelo19,Golosovsky12b,Golosovsky17,Okamura22}.
However, these studies did not investigate the temporal behaviour of $\sigma$, especially its consistent increase with time since publication.
Consequently, understanding the mechanisms underlying the $t^{H}$ law and developing a theoretical model that naturally reproduces this relationship remain important open problems, which we address below.

\section{Theoretical Framework\label{sec:theory}}

In this section, we present a fundamental theoretical framework for citation dynamics, incorporating the various spatiotemporal empirical observations and conditions discussed in the preceding sections.

\subsection{Heuristic considerations\label{sec:heuristic}}

In developing a theoretical framework for citation dynamics, we have already highlighted the necessity of incorporating stochastic elements.
But what form should this stochastic component take?
A close examination of the $t^{H}$ law strongly suggests that an extension of Brownian motion is a plausible candidate for two key reasons.

First, the $t^{H}$ law bears a striking resemblance to the well-known $\sqrt{t}$ law (with $H=\tfrac{1}{2}$) for SD growth in Brownian motion, a classic finding from Einstein's era.
Generalising the exponent from $\tfrac{1}{2}$ to $H\in(0,1)$ implies replacing Brownian motion with fractional Brownian motion \cite{Mandelbrot68,suppletext}, thereby extending the process via a Hurst parameter $H$.

Second, let us revisit the left-hand side of the $t^{H}$ law~\eqref{law:t^H} and refine the definition of the quantity $c$ with higher temporal resolution.
To this end, let $C(t)$ denote the cumulative citation count at time $t$, $\Delta t$ a sufficiently small time unit (one year in Fig.~\ref{fig:t^H}), and $\Delta C(t)=C(t+\Delta t)-C(t)$ the change over this interval.
The citation rate is then given by $c(t;\Delta t)=\Delta C(t)/\Delta t$.  
It is important to note that $C(t)$ does not represent the citation count of a single publication at time $t$ but rather the aggregate of discrete citation events reflecting attention to the collective body of publications.
Consequently, $C(t)$ is a monotonically increasing function that takes nonnegative integer values, forming a piecewise constant curve with numerous localised steps.
If the discrete jumps at each time $t=t_{k}$ are denoted by $a_{k}\in\mathbb{N}$ and $\Theta(x)$ represents the Heaviside step function, then $C(t)$ can be expressed as $C(t)=\sum_{k}a_{k}\Theta(t-t_{k})$.

Now consider the limit $\Delta t\to 0$.
At the discrete jump points in the graph of $C(t)$, the function is not differentiable in the usual sense; instead, its derivative behaves analogously to Dirac's $\delta$ function.
Specifically, the instantaneous attention at time $t$ can be expressed as:  
\begin{equation}
\lim_{\Delta t\to 0}c(t;\Delta t)=\frac{dC(t)}{dt}=\sum_{k}a_{k}\delta(t-t_{k})\,.\no
\end{equation}
If we refine the temporal resolution of attention history curves---typically observed as jump-and-decay patterns on an annual scale---to the level of instantaneous values and aggregate them across all relevant publications, then the resulting function, though continuous, appears highly nonsmooth.
This function represents a superposition of numerous sharp spikes, forming a randomly jagged trajectory reminiscent of Brownian motion.
In essence, the trajectory undergoes random fluctuations, while its overall trend follows a jump-and-decay pattern.

\subsection{The \textit{Ans\"{a}tze}\label{sec:ansatz}}

Building on the above mathematical and physical insights, the core \textit{Ans\"{a}tze} for modelling citation dynamics are as follows:
\begin{enumerate}[itemsep=0pt,label=(\textit{\roman*})]
\item\label{it:ans1} The citation count of a set of publications represents a discrete manifestation of the integral of a continuous, latent attention function, $X(t)$, over the publications' lifetime up to the present.
\item\label{it:ans2} The attention dynamics can be decomposed into two components: a deterministic trajectory and a stochastic component.  
Schematically,
\begin{equation}\label{ans:A*B}
X(t)=\mathcal{A}_\mathrm{deterministic}(t)\times\mathcal{B}_\mathrm{stochastic}(t)\,.
\end{equation}
Here $\mathcal{A}(t)$ represents the average or expected trajectory of $X(t)$, while the stochastic component $\mathcal{B}(t)$ captures random fluctuations.
\item\label{it:ans3} The latent attention exhibits cumulative advantage and adheres to Gibrat's law, meaning that its relative growth rate is independent of its current level.
\item\label{it:ans4} The stochastic component $\mathcal{B}(t)$ is driven by a fractional Brownian motion.
\end{enumerate}

Several remarks follow regarding these \textit{Ans\"{a}tze}.  
First, with respect to \textit{Ansatz}~\ref{it:ans1}, the relationship between the latent attention function $X(t)$ and its integral over the interval $[0,T]$, namely $C(T)$, will be discussed in detail later.
In particular, we will clarify how the continuous attention variable relates to the discrete citation count (see Eq.~\eqref{C<->X} and the discussion in Sec.~\ref{sec:conjecture}).

Regarding \textit{Ansatz}~\ref{it:ans2}, this paper does not impose any constraints on the specific functional form of the deterministic function $\mathcal{A}(t)$ or the mechanism that generates it.
Theoretically, it could take various forms.
Instead, our primary focus is on the stochastic function $\mathcal{B}(t)$.
We investigate its influence on attention dynamics, the distinct effects it induces, and the extent to which it serves as a fundamental driver of real-world citation patterns.

For \textit{Ansatz}~\ref{it:ans3}, we note a few points about the connection between Gibrat's law and preferential attachment.
While both involve multiplicative growth processes, there are some key differences.
Preferential attachment typically describes discrete nodes in a network acquiring new links, whereas Gibrat's law is usually applied to continuous variables, such as wealth or firm sizes---and here to latent attention $X(t)$.
Moreover, preferential attachment, though probabilistic at the microscopic level, effectively follows a global deterministic rule that yields a power law distribution if the attachment function is linear \cite{Barabasi99,*Barabasi02,*Jeong03,*Newman09}.
By contrast, Gibrat's law under stochastic fluctuations in the proportionality factor generates a log-normal distribution via geometric Brownian motion \cite{Sutton97,Okamura22}.
This feature will be crucial when we discuss the distribution of $C(T)$ in later sections.

Last, it is worth noting that the above set of \textit{Ans\"{a}tze} applies not only to citations but also to other forms of attention process.
In this context, citation can be replaced with similar indices that reflect the collective attention directed towards a given entity within a community.
See also the discussion in Sec.~\ref{sec:implication} \cite{suppletext}.

\subsection{Key properties of fractional Brownian motion\label{sec:fBm}}

Before incorporating fractional Brownian motion into our formalism based on \textit{Ansatz}~\ref{it:ans4}, we briefly summarise its definition and key properties.

\paragraph{Definition.}
Fractional Brownian motion, denoted as $\{B_{H}(t)\}_{t\geq 0}$ with the Hurst parameter $H\in (0,1)$, is a class of non-Markovian and nonmartingale, continuous-time Gaussian stochastic processes with stationary increments.
It satisfies
\begin{equation}
\Expt[B_{H}(t)]=B_{H}(0)=0\label{B_H:prop1}
\end{equation}
for all $t\geq 0$, i.e.\ it is a centred process, and the covariance is given by
\begin{equation}
\Expt[B_{H}(s)B_{H}(t)]=\frac{1}{2}\big(s^{2H}+t^{2H}-\abs{t-s}^{2H}\big)\label{B_H:prop2}
\end{equation}
for all $s,\,t\geq 0$.
Hence, in particular, $\Var[B_{H}(t)]=t^{2H}$.
This feature is useful when dealing with phenomena whose variance does not behave linearly in $t$, as in the present case with the citation process.
In the special case when $H=\tfrac{1}{2}$, one recovers the standard Brownian motion, $B_{1/2}(t)\equiv W(t)$, with covariance $\Expt[W(s)W(t)]=\tfrac{1}{2}\big(s+t-\abs{s-t}\big)=\min\{s,t\}$.

\paragraph{Memory effects.}
From Eqs.~\eqref{B_H:prop1} and \eqref{B_H:prop2}, it can be shown that for the fractional Brownian motion with $H\neq\tfrac{1}{2}$, a process after a given time $t$ depends not only on the situation at time $t$ but also on the entire past history of the process up to time $t$.
Specifically, the fractional Brownian motion increment $\Delta B_{H}(t,s)\coloneqq B_{H}(t)-B_{H}(s)$ with $0\leq s\leq t$ has the moment properties that $\Expt[\Delta B_{H}(t,s)]=0$ and $\Expt[\Delta B_{H}(t,s)^{2}]=\abs{t-s}^{2H}$, and therefore, the covariance of two nonoverlapping increments is given by 
\begin{equation*}
\Expt[\Delta B_{H}(t,s)\Delta B_{H}(s,0)]=\frac{1}{2}\big[t^{2H}-s^{2H}-(t-s)^{2H}\big]\,,
\end{equation*}
which is $\lesseqgtr 0$ if $H\lesseqgtr \tfrac{1}{2}$.

Consequently, if $H\in(\tfrac{1}{2},1)$, then the increments of fractional Brownian motion exhibit positive correlation (persistence), meaning a tendency for future increments to continue in the same direction as past increments, and the resulting motion is superdiffusive.
Conversely, if $H\in(0,\tfrac{1}{2})$, then the increments are negatively correlated (antipersistence), meaning a tendency for future increments to reverse the direction of past increments, producing subdiffusive motion.
Roughly speaking, the antipersistent case corresponds to more chaotic behaviour, while the persistent case reflects more structured and disciplined behaviour.

\paragraph{Fractality.}
The expression in Eq.~\eqref{B_H:prop2} also implies that fractional Brownian motion is a self-similar process \cite{Falconer14,Pipiras17}, i.e.\ $\{B_{H}(\lambda t)\}_{t\geq 0}\stackrel{\text{law}}{=}\{\lambda^{H}B_{H}(t)\}_{t\geq 0}$, obeying the same law for any $\lambda>0$.
The Hausdorff and box dimensions of the graph of $B_{H}(t)$, denoted $D$, is related to the Hurst parameter $H$ via the simple relation \cite[Theorem 16.8]{Falconer14}:
\begin{equation}\label{eq:fdim}
D=2-H\,.
\end{equation}
As can be seen from this formula, the closer the value of $H$ is to 1---that is, the stronger the persistent tendency---the trajectory approaches a smoother line. Conversely, the closer $H$ is to 0---that is, the stronger the antipersistent tendency---the trajectory becomes more jagged and complex in shape.

\paragraph{Applications.}
The distinctive and versatile nature of fractional Brownian motion makes it a compelling model for phenomena exhibiting both short-range and long-range dependence across various fields, including biology \cite{Allegrini98,*Jeon11}, hydrology \cite{Molz97}, telecommunications and network traffic \cite{Leland94,*Abry00,*Mikosch02}, and economics and finance \cite{Cutland95,*Comte98,Gatheral18}.
Indeed, numerous studies have demonstrated that many phenomena and challenges in these domains are better captured by fractional Brownian motion, which incorporates memory effects over time.

\subsection{The fractional stochastic differential equation\label{sec:fSDE}}

We now explain how fractional Brownian motion is explicitly incorporated into the formulation of citation dynamics.  
To begin, we express \textit{Ansatz}~\ref{it:ans1} in mathematical form.
Let $\Delta C(t_{i-1},t_{i})\in\mathbb{Z}_{\geq 0}$ be the number of citations a publication receives during the time interval $(t_{i-1},t_{i}]$ with $t_{0}=0$ and $\Delta t=t_{i}-t_{i-1}$, $i\in \mathbb{N}$.
For sufficiently small $\Delta t$ and with appropriate treatment, it can be approximated as $\Delta C(t_{i-1},t_{i})\approx X(t_{i})\dD t$, where $X(t)$ is the associated latent attention function at time $t$, assumed to be strictly positive for all $t>0$.
The cumulative citation count at time $t_{n}=n\dD t\equiv T$, denoted as $C(T)$, is then related to $X(t)$ by
\begin{equation}\label{C<->X}
C(T)\approx\sum_{i=1}^{n}X(t_{i})\dD t\approx\int_{0}^{T}X(t)\dd t
\end{equation}
for sufficiently small $\Delta t$ (or large $n$ for fixed $T$).
Note that while $C(0)=0$, it holds that $X(0)\equiv X_{0}>0$, which is treated as deterministic.

Now we explore the infinitesimal-time limit $\Delta t\to dt$.
To incorporate our set of \textit{Ans\"{a}tze}~\ref{it:ans1}--\ref{it:ans4}, we consider the following fractional SDE, which describes the evolution of latent attention:
\begin{equation}\label{fSDE}
dX(t)=X(t)\big[\alpha(t)\dd t+\beta\dd B_{H}(t)\big]\,.
\end{equation}
The time-dependent drift function $\alpha(t)$ determines the shape of the average attention history curve $X(t)$ [see Eq.~\eqref{X:Expt} below].
The parameter $\beta>0$, referred to as the volatility parameter---borrowing terminology from financial engineering---controls the magnitude of random fluctuations in $X(t)$ over the interval $dt$, reflecting external random events as discussed earlier.

As is evident, the fractional SDE~\eqref{fSDE} inherently incorporates our \textit{Ans\"{a}tze}~\ref{it:ans3} and \ref{it:ans4} by definition. 
In particular, \textit{Ansatz}~\ref{it:ans3} satisfies the property of nonlinear preferential attachment.
Compared to the SDE in the Black--Scholes model \cite{Black73,*Merton73}, this fractional SDE differs in two key aspects: Brownian motion is replaced by fractional Brownian motion, and the drift term $\alpha(t)$ is a function of time rather than a constant.
If $\alpha$ were constant, then this model would coincide with the fractional Black--Scholes model \cite{Hu03}.
In addition, compared to the SDE considered in Eqs.~(14) and (S3) in Ref.~\cite{Okamura22}, this fractional SDE differs in that Brownian motion is extended to fractional Brownian motion, while the volatility remains a time-independent constant.

Next we consider solving the fractional SDE~\eqref{fSDE}.
If $H=\tfrac{1}{2}$, then the process is a semimartingale, allowing the SDE to be solved using It\^{o} integration (Appendix C in Ref.~\cite{Okamura22}).
However, in the fractional SDE~\eqref{fSDE}, the process is nonsemimartingale, meaning that the integral $\int_{0}^{t}\dd B_{H}(s)$ cannot be evaluated in the It\^{o} sense.
Nonetheless, this integral is well-defined under the fractional Wick--It\^{o} integral \cite{Biagini04}.
The solution is:
\begin{equation}\label{fSDE_sol}
X(t)=X_{0}\exp\big[A(t)+\beta B_{H}(t)\big]\,,
\end{equation}
where the deterministic contribution in the exponent is given by
\begin{equation}\label{A_t}
A(t)\coloneqq\int_{0}^{t}\alpha(s)\dd s-\frac{\beta^{2}}{2}t^{2H}\,.
\end{equation}
The process described by Eq.~\eqref{fSDE_sol} is always positive and can approach zero only asymptotically.
This functional form also satisfies \textit{Ansatz}~\ref{it:ans2}, with $\mathcal{A}(t)\equiv X_{0}e^{A(t)}$ and $\mathcal{B}(t)\equiv e^{\beta B_{H}(t)}$.
From Eq.~\eqref{fSDE_sol}, one straightforwardly obtains:
\begin{align}
\Expt[X(t)]&=X_{0}\exp\int_{0}^{t}\alpha(s)\dd s\,,\label{X:Expt}\\
\Var[X(t)]&=\Expt[X(t)]^{2}\Big[\exp\big(\beta^{2}t^{2H}\big)-1\Big]\,,\label{X:Var}
\end{align}
and from Eq.~\eqref{X:Expt}, it follows that the average attention history curve, $u(t)\coloneqq\Expt[X(t)]$, relates to $\alpha(t)$ via
\begin{equation}\label{alpha<->u}
\alpha(t)=\frac{d\ln u(t)}{dt}\,.
\end{equation}

The key function in our analysis corresponds to what is termed the return in financial engineering, defined by
\begin{equation}\label{def:R}
R(t)\coloneqq \ln\ko{\frac{X(t)}{X_{0}}}=A(t)+\beta B_{H}(t)\,.
\end{equation}
Using the properties of fractional Brownian motion given by Eqs.~\eqref{B_H:prop1} and \eqref{B_H:prop2}, we obtain $\Expt[R(t)]=A(t)$ and $\Var[R(t)]=\beta^{2}\,t^{2H}$.
Hence, the $t^{H}$ law~\eqref{law:t^H} arises theoretically:
\begin{equation}\label{law_theory}
\SD[R(t)]=\beta t^{H}\,.
\end{equation}
Observe that while $\Expt[R(t)]$ depends on the time-dependent drift term $\alpha(t)$, $\SD[R(t)]$ is determined solely by the Hurst parameter $H$ and the constant volatility $\beta$, remaining independent of $\alpha(t)$.
Since $B_{H}(t)$ is Gaussian and $R(t)$ is a linear transformation of it [via Eq.~\eqref{def:R}], $R(t)$ itself follows a Gaussian distribution.
Thus, for any fixed $t$, $X(t)$ follows a log-normal distribution with mean $\ln X_{0}+\Expt[R(t)]$ and variance $\Var[R(t)]$.

\subsection{Interpolating between log-normal and power-law distributions\label{sec:model}}

The next task is to theoretically predict the distribution of $C(T)$, obtained via Eq.~\eqref{C<->X}, based on $X(t)$ given by Eq.~\eqref{fSDE_sol}.
As mentioned, $X(t)$ at each instant follows a log-normal distribution, yet its integral over time, $C(T)$, is not strictly log-normal in a rigorous sense, nor does it admit a simple closed form.
Nevertheless, under certain conditions it can be well approximated by a log-normal distribution, for instance through moment-matching or a Fenton--Wilkinson-type approximation \cite{Fenton60}.

Such conditions include the system being antipersistent, i.e.\ $H\in(0,\tfrac{1}{2})$, having a moderate $\beta$ value, and an $\alpha(t)$ that does not induce excessive fluctuations in the baseline trajectory of $X(t)$.
The last condition is empirically satisfied, as a typical average citation history curve follows a jump-and-decay pattern.  

To clarify, if we approximate $C(T)$ as $\sum_{i=1}^{n}X(t_{i})\dD t$ (i.e.\ treating it as a sum of random variables) and $u(t)$ has a damping effect that causes $X(t)$ to rapidly approach zero, then for sufficiently large $i$, the terms $X(t_{i})\dD t$ become negligible.
As a result, each individual term's contribution to the integral is suppressed, reducing the likelihood of heavy tails.  
Moreover, even when $X(t_{i})$ reaches a high value at a given instant, antipersistence naturally limits the occurrence of consecutive large peaks.
If $\beta$ is also moderate, then the probability of extreme peaks decreases further. 
Thus, under these conditions, the distribution of $C(T)$ tends to form a single, relatively compact `bump' with its peak near $C=0$ rather than an extended heavy tail.

In contrast, under what conditions is a power-law form more likely for $C(T)$?
This scenario is more probable if the system is persistent, i.e.\ $H\in(\tfrac{1}{2},1)$, $\beta$ is large, and $\alpha(t)$ ensures that $X(t)$ retains a finite, sustained contribution over a substantial time window.
Suppose that for some $i=i_{*}$, the term $X(t_{i_{*}})\dD t$ assumes an extremely large value due to a large $\beta$, thereby dominating the sum---this can be intuitively linked to jump discontinuities such as L\'{e}vy flights.
If the system also exhibits persistence, meaning large values tend to cluster, and if the range of $i$ over which these events may occur is sufficiently large, then the contribution of individual terms in the integral generates a heavy tail, making a power law a good approximation.

Accordingly, the conditions favouring a log-normal fit and those yielding a heavy-tailed (power-law) fit for $C(T)$ lie at opposite ends of this spectrum, with real-world citation distributions often interpolating between these extremes.

\section{Numerical Simulations\label{sec:simulation}}

Based on the theory we have developed, we perform numerical simulations of $X(t)$ and $C(T)$ to verify whether the theoretical predictions discussed in the previous section are reproduced and whether the observed macroscopic picture of citation dynamics is appropriately replicated.
We primarily focus on the antipersistent case, whereas the persistent case is discussed in the supplemental material \cite{suppletext}

\subsection{Simulation setup\label{sec:setup}}

For the numerical implementation of fractional Brownian motion, the \texttt{fbm} package \cite{Flynn19} in Python was used.
The Hurst parameter was set to $H=0.15$, ensuring a sufficiently strong antipersistent nature of the system. 
The fractional SDE~\eqref{fSDE} was discretised with a time step of $n=1{,}000$ and a final time of $T=10$, yielding $\Delta t=T/n=10^{-2}$.  
As a toy model for the function representing the average attention history curve  $u(t)$, the log-normal probability distribution function (PDF) was employed.
To ensure that $X_{0}>0$, we used a rescaled version, given by
\begin{equation*}
\hat{u}(t)=\frac{X_{0}t_{0}}{t+t_{0}}\exp\kko{-\frac{(\ln (t+t_{0})-\theta)^{2}}{2\omega^{2}}+\frac{(\ln t_{0}-\theta)^{2}}{2\omega^{2}}}\,,
\end{equation*}
which implies that, for the associated time-dependent drift term, in view of Eq.~\eqref{alpha<->u},
\begin{equation}\label{alpha_hat}
\hat{\alpha}(t)=\frac{\theta-\omega^{2}-\ln (t+t_{0})}{\omega^{2}(t+t_{0})}\,.
\end{equation}
Here $\omega$ is the shape parameter, $e^{\theta}$ represents the median of the log-normal PDF, and $t_{0}>0$ is the location parameter.

Additionally, we considered five cases for the volatility parameter, setting $\beta=0.2$, $0.5$, $0.8$, $1.0$ and $1.5$, to assess the impact of volatility variations on the system's temporal evolution.  
For the parameters describing Eq.~\eqref{alpha_hat}, we adopted $t_{0}=10^{-50}$, $\theta=0.48$ and $\omega=0.8$.  
The value of $X_{0}$ was determined so that the computed values of $\{\hat{X}_{i}\equiv\hat{X}(t_{i})\}_{i=1}^{n}$ remained within a realistic range for citation counts, accounting for the interplay with other parameters.  
Specifically, we set $X_{0}=10^{-53}$.

\subsection{Simulating attention curves\label{sec:trajectory}}

\begin{figure*}[t!]
\centering
\begin{minipage}[t]{0.48\textwidth}
\includegraphics[width=0.92\textwidth]{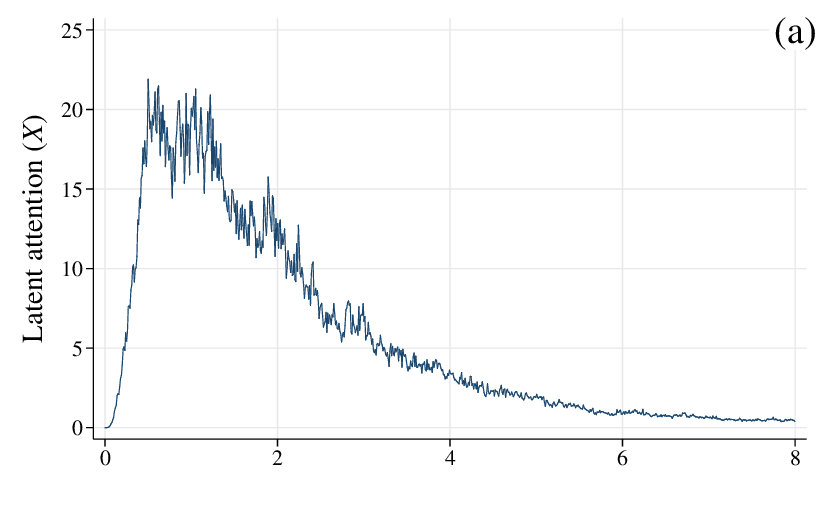}\\
\includegraphics[width=0.92\textwidth]{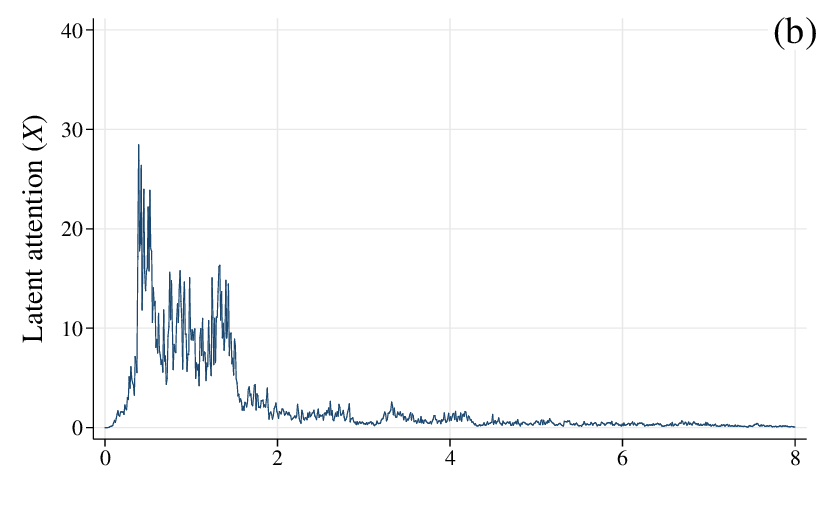}\\
\includegraphics[width=0.92\textwidth]{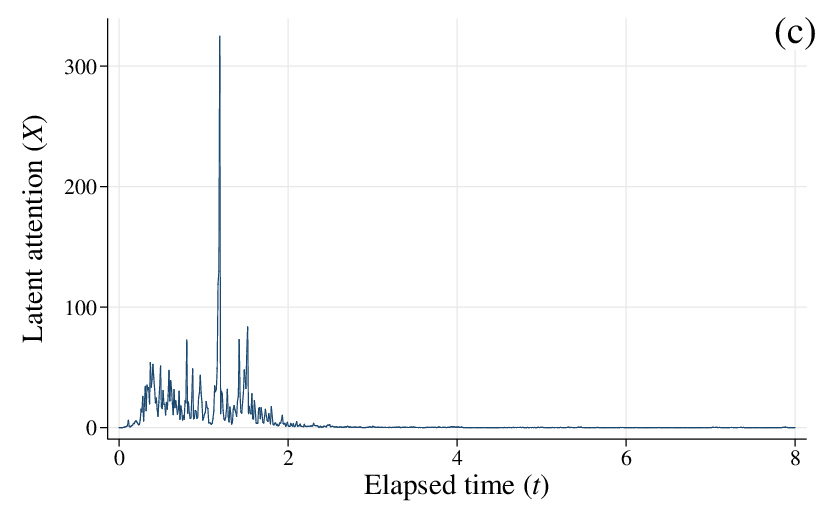}
\caption{Simulation results of the attention curve $\hat{X}(t)$ for an antipersistent system ($H=0.15$).  
The parameters in $\hat{\alpha}(t)$ are set to $\theta=0.48$ and $\omega=0.8$.  
The volatility values are (a) $\beta=0.2$, (b) $\beta=0.8$ and (c) $\beta=1.5$.}
\label{fig:antipers_path}
\end{minipage}
\hspace{1.5em}
\begin{minipage}[t]{0.48\textwidth}
\includegraphics[width=0.92\textwidth]{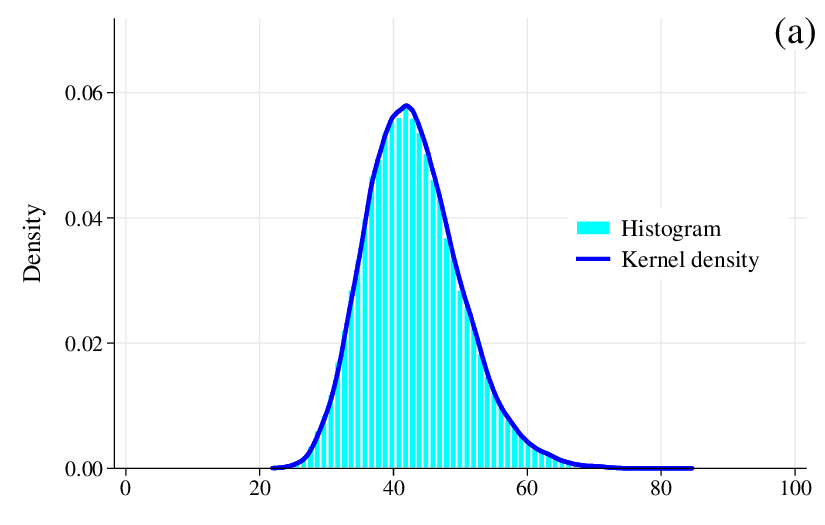}\\
\includegraphics[width=0.92\textwidth]{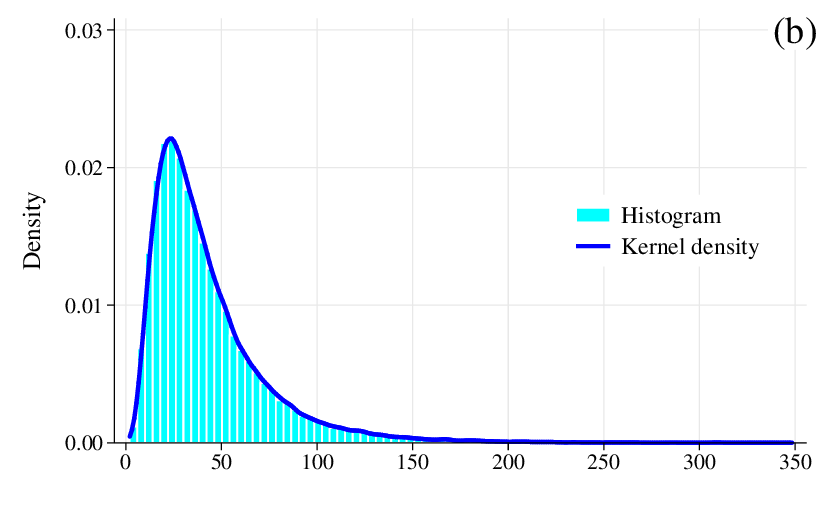}\\
\includegraphics[width=0.92\textwidth]{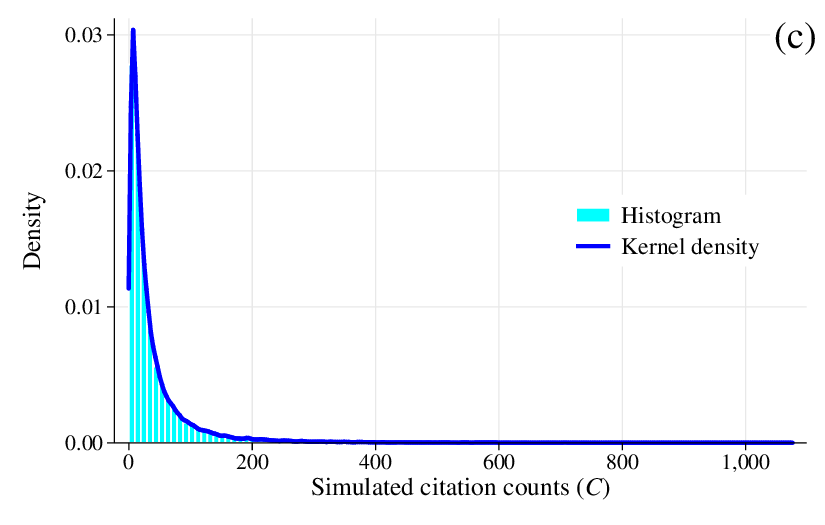}
\caption{Monte Carlo simulation results for the distribution of $\hat{C}(T)$, generated using the same parameter settings as in Fig.~\ref{fig:antipers_path}, with $N_\mathrm{s}=50{,}000$ trials.  
The volatility values are (a) $\beta=0.2$, (b) $\beta=0.8$ and (c) $\beta=1.5$.}
\label{fig:antipers_histo}
\end{minipage}
\end{figure*}

With the above implementation and parameter settings, the numerical simulation results for the trajectory of $\{\hat{X}_{i}\}_{i=1}^{n}$ are shown in Fig.~\ref{fig:antipers_path}.  
For $\beta=0.2$ [Fig.~\ref{fig:antipers_path}(a)], the trajectory closely follows a log-normal PDF, as theoretically expected.
As $\beta$ increases [Figs.~\ref{fig:antipers_path}(a) to \ref{fig:antipers_path}(b) to \ref{fig:antipers_path}(c)], fluctuations become more pronounced, and the trajectory exhibits greater instability, aligning with theoretical predictions.
Furthermore, due to the decay effect of $\hat{u}(t)$, the period during which the attention undergoes large fluctuations remains relatively short.
Beyond a certain point, only negligible fluctuations persist, further reinforcing theoretical expectations.  

Additionally, an increase in volatility raises the likelihood of rare but extreme values of $\hat{X}$; for example, note the high peak of $\hat{X}>300$ in Fig.~\ref{fig:antipers_path}(c).
However, a crucial aspect of the current setup---where $H$ is significantly less than $0.5$, indicating a strongly antipersistent system---is that such peaks do not persist or occur repeatedly over extended periods.
Instead, they appear as relatively isolated and sporadic events.  
In other words, when $\hat{X}$ reaches a large value at a given moment, it is likely to decrease soon after due to a counteracting effect.
Consequently, contributions to the integral become more dispersed, reducing the tendency for the distribution of $\hat{C}(T)$ to develop an excessively heavy tail.

\subsection{Monte Carlo simulation of citation distributions\label{sec:MonteCarlo}}

Under the same settings described above, we performed $N_\mathrm{s}=50{,}000$ Monte Carlo simulations to examine the distribution of cumulative citation counts up to time $T$.  
Here $N_\mathrm{s}$ can be simply interpreted as the number of publications in the similated world.
The simulated citation count was obtained as follows \cite{suppletext}:
\begin{equation}\label{def:C_tilde}
\hat{C}(T)\coloneqq\ceil*{\sum_{i=1}^{n}\hat{X}_{i}\dD t}\,,
\end{equation}
where $\lceil\cdot\rceil$ denotes the ceiling function.
The resulting histogram is shown in Fig.~\ref{fig:antipers_histo}.
As volatility increases [Fig.~\ref{fig:antipers_histo}(a) to \ref{fig:antipers_histo}(b) to \ref{fig:antipers_histo}(c)], the reproduced distribution becomes closer to the highly skewed distributions commonly observed empirically.
Moreover, given the present settings for $H$ and $\hat{\alpha}(t)$, $\hat{C}(T)$ tends not to develop a heavy tail for any choice of $\beta$, making it more amenable to approximation by a single log-normal distribution.

\begin{figure*}[t!]
\centering
\includegraphics[width=0.49\linewidth]{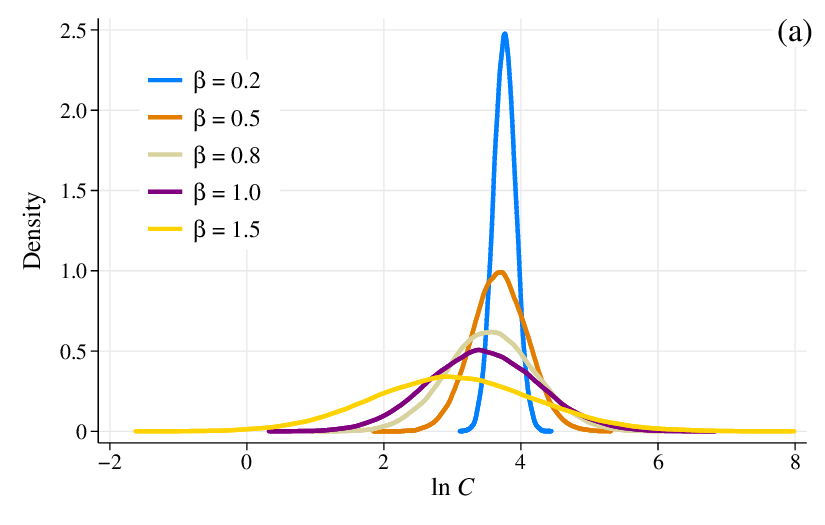}
\hspace{0.5em}
\includegraphics[width=0.49\linewidth]{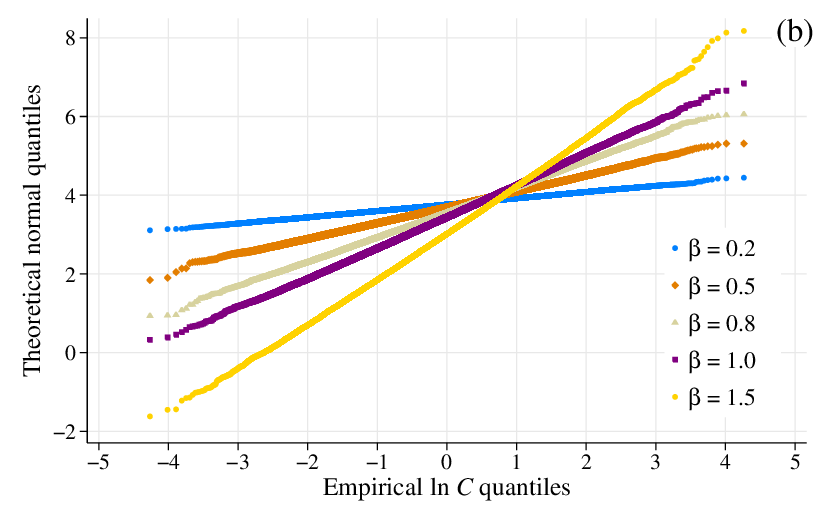}
\caption{(a) Kernel density estimation of the distribution of $\ln\hat{C}(T)$ using the same data as in Fig.~\ref{fig:antipers_histo}, overlaid with different values of volatility ($\beta$) for visual clarity.  
(b) Q--Q plot of the same $\ln\hat{C}(T)$ distribution against theoretical normal quantiles.}
\label{fig:antipers_QQ}
\end{figure*}
\begin{table}[b!]
\caption{Summary of the estimated shape parameter ($\hat{\sigma}$) for the fitted log-normal distribution under varying volatility parameters ($\beta$), given $H=0.15$ (antipersistent) and $\hat{\alpha}(t)$ from Eq.~\eqref{alpha_hat} with $\theta=0.48$ and $\omega=0.8$.  
All cases satisfy $p<0.001$.}
\label{tab:antipers_QQ}
\setlength{\tabcolsep}{8pt}
\vspace{0.5em}
\begin{tabular}{cc}
\midrule\addlinespace[-0.1ex]
\midrule\\[-1.7em]
$\beta$ & $\hat{\sigma}$ \\[-0.2em]
\midrule
0.2 & $0.16 \pm 9.1\times 10^{-6}$ \\
0.5 & $0.40 \pm 2.6\times 10^{-5}$ \\
0.8 & $0.64 \pm 7.2\times 10^{-5}$ \\
1.0 & $0.80 \pm 1.0\times 10^{-4}$ \\
1.5 & $1.19 \pm 2.2\times 10^{-4}$ \\
2.0 & $1.57 \pm 3.5\times 10^{-4}$ \\
\midrule\addlinespace[-0.1ex]
\midrule
\end{tabular}
\end{table}

The kernel density functions of the distribution of $\ln\hat{C}(T)$ for various values of $\beta$ are shown in Fig.~\ref{fig:antipers_QQ}(a).
Additionally, to assess its log-normality, a quantile--quantile (Q--Q) plot of the same $\ln\hat{C}(T)$ distribution against theoretical normal quantiles is presented in Fig.~\ref{fig:antipers_QQ}(b).
The linearity of this plot suggests that $\hat{C}$ closely follows a log-normal distribution.  
The verification of this linearity using the least-squares method is summarised in Table~\ref{tab:antipers_QQ}, which clearly shows that the shape parameter $\hat{\sigma}$ of the fitted log-normal distribution increases as $\beta$ increases.
According to previous studies \cite{Radicchi08,*Evans12,*Chatterjee16,*DAngelo19,Golosovsky12b,Golosovsky17,Okamura22}, the corresponding shape parameter typically falls within the range of approximately 1 to 1.4.  
This scenario corresponds to setting $\beta$ in the range of roughly 1.5--2, precisely when the distribution of $\hat{C}$ becomes markedly skewed, as observed in Fig.~\ref{fig:antipers_histo}(c).

For comparison, it is instructive to conduct a hypothetical simulation in which the system exhibits a persistent memory effect and attention does not decay \cite{suppletext}.  
In this scenario, compared to the antipersistent case, the trajectory of attention becomes smoother, with less pronounced jagged fluctuations (Fig.~\ref{fig:pers_path}).
The distribution of $\hat{C}(T)$ develops a fatter tail, and this tendency becomes more pronounced as $\beta$ increases [Fig.~\ref{fig:pers_histo}(a) to \ref{fig:pers_histo}(b) to \ref{fig:pers_histo}(c)].
Moreover, the high-citation region can be well approximated by a shifted power law [Fig.~\ref{fig:pers_QQ}(a)--\ref{fig:pers_QQ}(c)].  
All these findings align with the theoretical predictions presented in the previous section.

\subsection{Simulating the $\bmk{t^{H}}$ law\label{sec:simu_t^H}}

\begin{figure}[t!]
\centering
\includegraphics[width=1.0\linewidth]{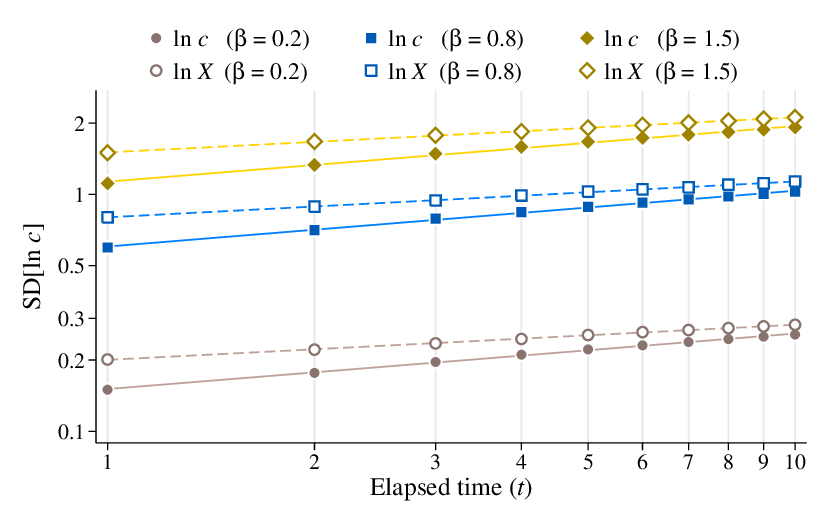}
\caption{Scatter plot of $(\tau_{k},\hat{c}_{k})$ (filled markers) and $(\tau_{k},\hat{X}(\tau_{k}))$ (unfilled markers) from the same data as in Fig.~\ref{fig:antipers_histo}, for $k=1\text{--}10$, along with their linear regression fits.  
In both cases, the error bars are negligibly small and are therefore omitted for clarity.}
\label{fig:verify_law}
\end{figure}

Finally, we confirm that the Monte Carlo simulations successfully replicate the $t^{H}$ law observed in Fig.~\ref{fig:t^H}.  
We used the same settings as in the earlier antipersistent scenario, except for assigning $X_{0}=1$.
For the `yearly' time intervals in Fig.~\ref{fig:t^H}, we set $\tau_{k}=k\in\mathbb{N}$.  
Then, for each period $\{(\tau_{k-1},\tau_{k}]\}_{k=1}^{10}$, we compute the number of citations acquired during year $k$, denoted as $\hat{c}_{k}$, via the same method as Eq.~\eqref{def:C_tilde}:
\begin{equation}\label{def:c_k}
\hat{c}_{k}\coloneqq \ceil*{\sum_{t_{i}\in(\tau_{k-1},\tau_{k}]} X_{i}\dD t}\,.
\end{equation}
Repeating this procedure $N_\mathrm{s}$ times yields $\{\SD[\ln\hat{c}_{k}]\}_{k=1}^{10}$.  
Figure~\ref{fig:verify_law} shows the results for three cases: $\beta=0.2$, $0.8$ and $1.5$ (filled markers).  
Examining the relationship between $\ln\SD[\ln \hat{c}_{k}]$ and $\ln\tau_{k}$ reveals a linear trend, confirming that the simulation reproduces the $t^{H}$ law seen in Fig.~\ref{fig:t^H}.

An important caveat here is that the values of $\hat{H}$ and $\hat{\beta}$ estimated from linear regression on this scatter plot systematically deviate by an approximately constant factor from the original $H$ and $\beta$ set in the simulation.  
To validate this point, we also simulated $X(t)$ at specific time points $t=\tau_{k}$ and analysed the relationship between $\ln\SD[\ln\hat{X}(\tau_{k})]$ and $\ln\tau_{k}$.
The linear regression analysis of this relationship precisely reproduces the original parameter values, as predicted by theory, Eq.~\eqref{law_theory}, and is shown by the unfilled markers in Fig.~\ref{fig:verify_law}.
The primary reason for this discrepancy is that the coarse summation operation in Eq.~\eqref{def:c_k} smooths out local fluctuations in $X(t)$.
As a result, $H$ tends to be overestimated, while $\beta$ tends to be underestimated.
The same caution must be exercised when performing empirical analysis on arXiv data in the subsequent section.

\section{Empirical Data Analysis\label{sec:analysis}}

In the previous section, we confirmed that simulations based on the fractional SDE~\eqref{fSDE} accurately reproduced our theoretical predictions, reinforcing the validity of our framework.
In this section, we estimate $H$ and $\beta$ using empirical citation data from arXiv e-prints, applying our model to a real-world system.

\subsection{Exploring citation dynamics in arXiv\label{sec:arXiv_data}}

We analysed the relationship between the number of citations received annually and the time elapsed since posting for arXiv e-prints published between 1995 and 2006, tracking citations up to February 2020.
The dataset used is the same as that employed in Refs.~\cite{Okamura21,Okamura22}.
Note that the arXiv dataset includes papers that were subsequently published in peer-reviewed journals, and that the citation data obtained from Semantic Scholar \cite{semanticscholar} include citations both from and to peer-reviewed publications.

Since these citation data are aggregated on an annual basis, the number of available data points is at most a few dozen, which is substantially smaller than the long-term datasets typically employed in financial engineering \cite{suppletext}.
Because of this limitation, we cannot effectively apply widely used techniques such as rescaled range analysis (R/S) \cite{Hurst51,Mandelbrot72} or detrended fluctuation analysis (DFA) \cite{Peng94,*Hu01,*Chen02}.
Under such constraints, the $t^{H}$ law presented in Eq.~\eqref{law_theory} serves as a key methodological guideline for estimating the fundamental parameters of the system.
By leveraging this relation, we determine the values of $H$ and $\beta$ through a straightforward regression analysis.

As discussed at the end of the previous section, this estimation method introduces a systematic bias that depends on how coarse the annual time resolution is relative to the scale of system fluctuations.
For instance, the estimate of $H$ obtained in Fig.~\ref{fig:t^H} can reasonably be interpreted as an overestimate.
Recognising this limitation as an inherent constraint of the dataset, we adopt a conservative approach: the estimated $\hat{H}$ is treated as an implied upper bound for $H$, while $\hat{\beta}$ is considered an implied lower bound for $\beta$.

For each individual e-print in the dataset, we compiled citation data annually, defining the elapsed years as $\tau_{k}=k\in\mathbb{N}$ with $\Delta\tau=\tau_{k}-\tau_{k-1}=1$.
Specifically, we constructed the dataset $\{\Delta C_{k}\}$, representing the number of citations received in year $k$, i.e., $k$ years since publication, and then computed $\big\{R_{k}\coloneqq\ln\big(\Delta C_{k}/\Delta\tau\big)\big\}$  in accordance with Eq.~\eqref{def:R}.
Here citation records with $\Delta C_{k}=0$ were excluded from the analysis.
Finally, for each year, we computed $\{\SD[R_{k}]\}$.
Performing a linear regression on the obtained data points $\mathrm{P}_{k} \equiv (\ln\tau_{k},\ln\SD[R_{k}])$, we estimated the upper bound of $H$ from the slope. 
Similarly, the lower bound of $\beta$ was obtained by exponentiating the intercept of the regression model.  
It is important to note that $\SD[R_{k}]$ in this regression represents the SD of citation counts during year $k$, which is distinct from the SD of the logarithm of cumulative citations over $[0,T]$, denoted as $\sigma$ earlier.

Figure~\ref{fig:arXiv_allyear} presents scatter plots of $\{\mathrm{P}_{k}\}$ for each publication year, overlaid with regression lines, while Table~\ref{tab:H-beta_hat} summarises the corresponding linear regression results.  
For all groups of e-prints published in different years, the $t^{H}$ law holds well, with the results for 2001 matching those in Fig.~\ref{fig:t^H}.
The estimated values $\hat{H}$ and $\hat{\beta}$ over time are shown in Figs.~\ref{fig:arXiv_allfield}(a) and \ref{fig:arXiv_allfield}(b), respectively.
These remain approximately within the ranges of $0.08$ to $0.15$ for $\hat{H}$, with an average of $\hat{H} \approx 0.13$, and $0.53$ to $0.65$ for $\hat{\beta}$, without any noticeable deviations or trends contradicting constancy.

The observed $t^{H}$ law can be interpreted as follows.
Most papers receive no citations from the outset or eventually fall to $\Delta C_{k}=0$, forming the sharp peak at zero in the distribution.
A small subset of long-lived papers maintain $\Delta C_{k}>0$, exhibiting diverse citation trajectories.
Consequently, the distribution becomes a mixture: a largely static mass at zero and a gradually widening tail of nonzero values.
The $t^{H}$ law reflects the increasing heterogeneity in $\ln(\Delta C_{k})$ within this tail over time.

Additionally, performing the same estimation separately for different arXiv subject categories reveals notable variations in these estimated values across disciplines (Fig.~\ref{fig:arXiv_byfield}).
These differences likely reflect domain-specific traits, such as typical patterns of scientific discovery, citation behaviours within each community, and publishing norms inherent to each field \cite{suppletext}.

It should be noted that these results are obtained through a simplified estimation procedure and based on data with a limited number of points, and therefore must be interpreted with the aforementioned biases in mind.
Nevertheless, they offer statistically solid and qualitatively robust insights into memory effects in citation dynamics.
From these estimations, we conclude that the citation dynamics of e-prints on arXiv follow a strongly antipersistent stochastic process.
Combining this with the simulation insights from Sec.~\ref{sec:simulation} strengthens the theoretical basis for why the cumulative citation distribution of arXiv e-prints (as observed in Ref.~\cite{Okamura22}) is well approximated by a highly skewed log-normal distribution with a pronounced peak near zero citations.

\subsection{Hidden fractality in citation trajectory\label{sec:fractality}}

Using the previously estimated upper bound of the Hurst parameter, we can apply Eq.~\eqref{eq:fdim} to estimate the lower bound of the fractal (Hausdorff) dimension of the fractional Brownian motion trajectories that underlie the arXiv system under consideration.
The result is $\hat{D}=2-\hat{H}\approx 1.87$.  

Fractal phenomena in human social activities have been explored in various contexts, including urban morphology and city structures \cite{Batty94}, financial markets \cite{Mandelbrot72}, and complex networks \cite{Song05,*Song06}.
However, quantitatively estimating fractal dimensions in such systems remains challenging.
In citation dynamics, although the above finding is indirect rather than based on direct observation, it is noteworthy that it provides quantitative insight into the underlying intrinsic fractal structure.

\section{Discussion\label{sec:discussion}}

We now examine the interpretation of the results obtained, compare them with previous studies, discuss broader implications, identify unresolved issues, and explore potential directions for future research, including conjectures.

\subsection{Philosophical and theoretical implications\label{sec:implication}}

The theoretical framework of citation network growth dynamics presented in this paper, based on the fractional SDE~\eqref{fSDE}, establishes a microscopic foundation for previously recognised insights, which were largely derived from macroscopic observations, such as the distribution of cumulative citations and characteristic citation history curves.  
This approach parallels the role of statistical mechanics in thermodynamics, which seeks to explain empirical laws by uncovering the underlying microscopic mechanisms through dynamical equations.  
In the context of citation dynamics, this framework integrates mathematical physics perspectives with traditional bibliometric and scientometric methodologies.  
By incorporating the `science of science' approach \cite{Fortunato18}, it serves as a bridge between these disciplines, providing a more rigorous theoretical foundation for observed citation phenomena.

Furthermore, these findings hold significant implications for real-world applications of fractional Brownian motion.
It is well known that its mathematical structure fundamentally differs depending on whether the Hurst parameter is greater or less than $\tfrac{1}{2}$ \cite{Mandelbrot68,Biagini08}.  
Historically, the regime $H\in(\tfrac{1}{2},1)$ has been more extensively studied, particularly in finance, due to the long-range dependence observed in certain market data.
In contrast, the case $H\in(0,\tfrac{1}{2})$, where citation dynamics fall, has received comparatively less attention in real-world applications.
However, models exhibiting antipersistence in this regime have recently gained significant attention in finance, particularly in the study of rough volatility \cite{Gatheral18}.

The theoretical framework developed in this study applies broadly to both antipersistent and persistent systems.
Given the universality of the adopted \textit{Ans\"{a}tze}~\ref{it:ans1}--\ref{it:ans4}, it extends beyond conventional citation processes and may serve as a foundational theory for analogous attention dynamics \cite{Candia19,*Lorenz-Spreen19}.
In the empirical analysis in Sec.~\ref{sec:analysis}, we examined citation as a well-defined example of attention, focusing on annual citation counts.
If the theory applies more broadly, then other forms of attention should also adhere to the $t^{H}$ law when the SD of the logarithm of attention at each point are analysed.
Moreover, cumulative attention is generally well approximated by a log-normal distribution but may display power-law behaviour depending on volatility, memory effects, and the shape of the average attention history curve \cite{suppletext}.
These insights indicate diverse applications across multiple fields, highlighting the need for further research to refine and validate the theory's broader relevance.

\subsection{Memory effects and non-Markovianity\label{sec:memory}}

Numerous studies have made significant contributions to modelling and understanding citation dynamics using non-Markovian stochastic processes \cite{Glanzel95b,*Klemm02,Hajra06,Eom11,Golosovsky12b,Golosovsky12,Wang13,Golosovsky17}. 
These approaches have been particularly effective in capturing key features of citation growth by incorporating time-dependent weighting functions or memory kernels into the citation acquisition process (i.e.\ the link generation process in network growth models) to account for `ageing' or `obsolescence' effects.  
In such models, past history influences future evolution in a deterministic, time-dependent manner.
While this offers valuable insights, relying solely on this mechanism to introduce non-Markovian characteristics does not fully account for the empirical observations, including the $t^{H}$ law~\eqref{law:t^H}.

In contrast, the approach presented in this paper inherently incorporates memory effects through the stochastic fluctuations of fractional Brownian motion.
As a result, the $t^{H}$ law naturally emerges from our theoretical prediction in Eq.~\eqref{law_theory} and is empirically validated through simulations (Fig.~\ref{fig:verify_law}).
However, the presence of memory effects in stochastic fluctuations does not preclude the possibility that other forms of memory---potentially arising from different stochastic processes---may also contribute to the observed phenomena.  
Moreover, memory effects may reside not only in the stochastic component ${\mathcal B}(t)$ of Eq.~\eqref{ans:A*B} but also in the deterministic component ${\mathcal A}(t)$.
By selectively integrating elements of previously proposed models where appropriate, a more comprehensive framework with enhanced explanatory and predictive capabilities could potentially be developed.

Further progress may also be achieved by refining how stochastic fluctuations are introduced.  
Potential directions include employing mixed fractional Brownian motions \cite{Cheridito01,*Thale09} or reset geometric fractional Brownian motion \cite{Stojkoski21}.
For instance, given the interdisciplinary variations in the estimated $\hat{H}$ and $\hat{\beta}$ values observed in Sec.~\ref{sec:analysis} (Fig.~\ref{fig:arXiv_byfield}), it is plausible that citation dynamics across the broader academic landscape, including the arXiv disciplines used here as a testbed, might be better described by a mixed fractional Brownian motion.  
Such a process can be expressed as either a discrete sum, $Z(t)=\sum_{\ell\in\mathcal{S}}\beta_{\ell}B_{H_{\ell}}(t)$, or a continuous analogue, $Z(t)=\int_{\mathcal{S}}\beta(\rho)B_{H(\rho)}(t)\dd\rho$.  
Here $B_{H_{\ell}}(t)$ are independent fractional Brownian motions with $H_{\ell}\in(0,1)$ and $\beta_{\ell}>0$, where $\mathcal{S}$ denotes a proper summation domain, with a similar formulation applying to the continuous version.

\subsection{Unanswered questions and conjecture\label{sec:conjecture}}

Despite the findings presented here, several fundamental questions remain open.  
Below, we outline some of these issues, each representing a compelling direction for future research.

First, it is unclear whether the fractional Brownian motion governing attention dynamics can be derived from first principles.  
We introduced the idea that citation processes are driven by fractional Brownian motion mainly based on empirical observations, validating its plausibility through simulations and real-world data.  
However, our construction of the fractional SDE was not derived from first principles.  
A rigorous explanation for why fractional Brownian motion should naturally emerge in this context remains lacking.

Second, the interpretation and justification of the relationship between the attention function and the citation counts expressed in Eq.~\eqref{C<->X} remain unresolved.  
What exactly does this equation signify, and how can it be justified rigorously?  
The meaning of the approximation `$\approx$' also remains unclear under the current discussions.  
When associating the integral of the continuous variable $X(t)$ from our model with the discrete citation count $C(T)$ in reality, what mathematical operations or conceptual interpretations underlie this correspondence?  
In our simulations, we used the ceiling function in Eq.~\eqref{def:C_tilde} as a convenient discretisation device, but its role is merely pragmatic rather than theoretically fundamental.

Third, and closely related, is the question of what physical entity the latent attention function $X(t)$ represents.  
In our formulation, it was introduced as an auxiliary mathematical device for describing the statistical distribution of $C(T)$ in a network growth model.  
By hypothesising a fractional SDE and solving it, we reproduced various empirical features of $C(T)$.  
Yet it remains unclear whether $X(t)$ corresponds to any directly measurable quantity in the real world.

One possible resolution to these questions is suggested by classical mathematical arguments based on Donsker's invariance principle \cite{Donsker51,*Taqqu75,*Enriquez04}.  
Drawing on this theory, we propose the following conjecture: beneath citation processes, which are inherently discrete, there exists an underlying integer-valued stochastic process.  
By repeating this process many times and applying appropriate normalisation, the functional central limit theorem ensures the emergence of a discrete centred Gaussian process.
Taking the scaling limit of this discrete process then naturally leads to fractional Brownian motion as a characteristic feature of the system.

\section{Summary and Conclusion\label{sec:summary}}

In this study, we developed a model that is consistent with both previously established empirical findings and a newly identified empirical law by capturing the stochastic nature and dynamics of citation behaviour through an SDE incorporating random fluctuations.

Specifically, we adopted \textit{Ans\"{a}tze}~\ref{it:ans1}--\ref{it:ans4} (Sec.~\ref{sec:preliminary}) and formulated the fractional SDE~\eqref{fSDE} that governs latent attention, whose discrete realisation corresponds to citation counts.
By linking its solution, Eq.~\eqref{fSDE_sol}, to citation dynamics via Eq.~\eqref{C<->X}, we demonstrated---both theoretically (Sec.~\ref{sec:theory}) and through numerical simulations (Sec.~\ref{sec:simulation})---that the resulting model satisfies four key properties:
\begin{enumerate}[itemsep=0pt]
\item It permits a variety of citation history curve patterns.
\item It adheres to the principle of cumulative advantage, specifically satisfying nonlinear preferential attachment.
\item The SD of the logarithm of citations per unit time follows the $t^{H}$ law~\eqref{law:t^H} with respect to the time elapsed since publication.
\item Depending on the system conditions, the citation distribution may be well approximated by a log-normal distribution or, particularly in the high-degree regime, by a heavy-tailed power law.
\end{enumerate}  
Furthermore, applying our validated model to empirical data on arXiv e-prints has enhanced the microscopic understanding of citation phenomena observed in real-world settings (Sec.~\ref{sec:analysis}).  
Additionally, we have successfully estimated the fractal dimension associated with the system's attention.

Thus, the in-degree distribution of citation networks, a long-standing subject of research, is naturally explained within our theoretical framework based on a few key system characteristics: the Hurst parameter ($H$), the volatility parameter ($\beta$), and the time dependence of the deterministic component [$\alpha(t)$ or $A(t)$].
A central aspect of this model is the use of fractional Brownian motion to characterise the system's stochastic component, with the memory effect governed by the Hurst parameter $H$.

Specifically, we demonstrated that when $H\in(0,\tfrac{1}{2})$, indicating an antipersistent nature, and this is combined with the empirical observation that the average citation history curve decays over time, the model effectively accounts for why citation distributions are often well approximated by log-normal distributions.  
Conversely, when fluctuations occasionally lead to sustained high levels of attention, the citation distribution is better approximated by a power law.
This distinction between the two regimes offers a coherent explanation for the empirical variability observed in citation dynamics.

As discussed in Sec.~\ref{sec:discussion}, this perspective also gives rise to new questions from physical, mathematical, and statistical viewpoints.  
Addressing these open problems is expected to further advance our understanding of network dynamics and foster future developments in nonlinear dynamics, nonequilibrium statistical mechanics, as well as chaos and fractal theory.

\begin{acknowledgments}
The author would like to thank the anonymous referees of \textit{Phys.~Rev.~E} for their valuable comments on the manuscript.
The views and conclusions expressed herein are solely those of the author and should not be construed as necessarily reflecting the official policies or endorsements---whether explicit or implicit---of any organisations with which the author is presently or has previously been affiliated.
\end{acknowledgments}

\section*{Data Availability}
The data that support the findings of this article are openly available \cite{Okamura21,Okamura25z}.


%


\onecolumngrid
\clearpage

\renewcommand{\theequation}{S\arabic{equation}}
\renewcommand{\thefigure}{S\arabic{figure}}
\renewcommand{\thetable}{S\arabic{table}}
\setcounter{equation}{0}
\setcounter{figure}{0}
\setcounter{table}{0}
\setcounter{page}{1}

\vspace*{0em}
\begin{center}
\textbf{\fontsize{13pt}{14pt}\selectfont
          Supplementary Materials          %
}\\[1em]
{\large for {`\mtitle'}}
\end{center}
\vspace{2.0em}

\thispagestyle{suppl1}
\pagestyle{suppl2}

\noindent
This Supplemental Material accompanies the article published in \textit{Phys.~Rev.~E} \textbf{112}, 044304 (2025), DOI:
\href{https://doi.org/10.1103/l2xd-43n9}{10.1103/l2xd-43n9}.

\begin{figure}[h!]
\centering
\begin{minipage}[t]{0.48\textwidth}
\includegraphics[width=0.87\textwidth]{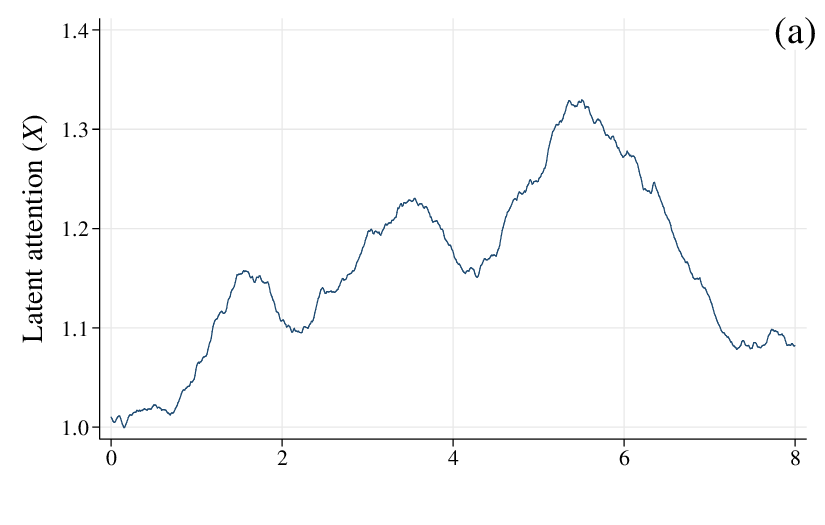}\\
\includegraphics[width=0.87\textwidth]{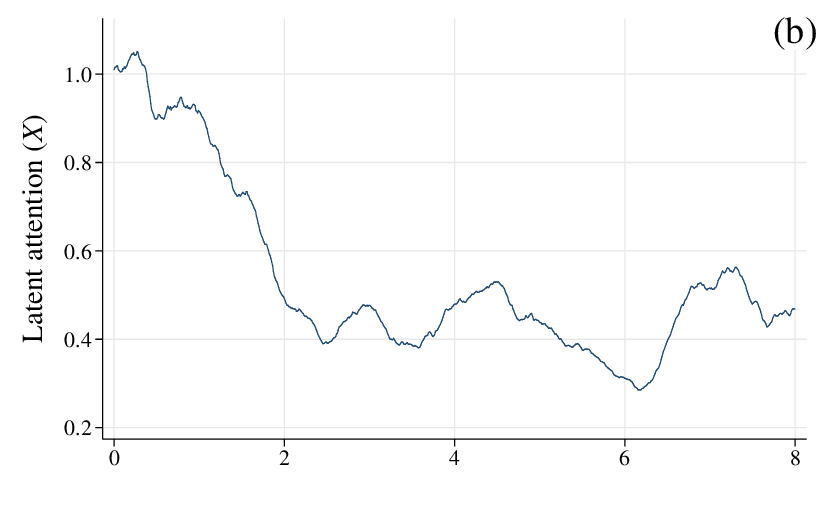}\\
\includegraphics[width=0.87\textwidth]{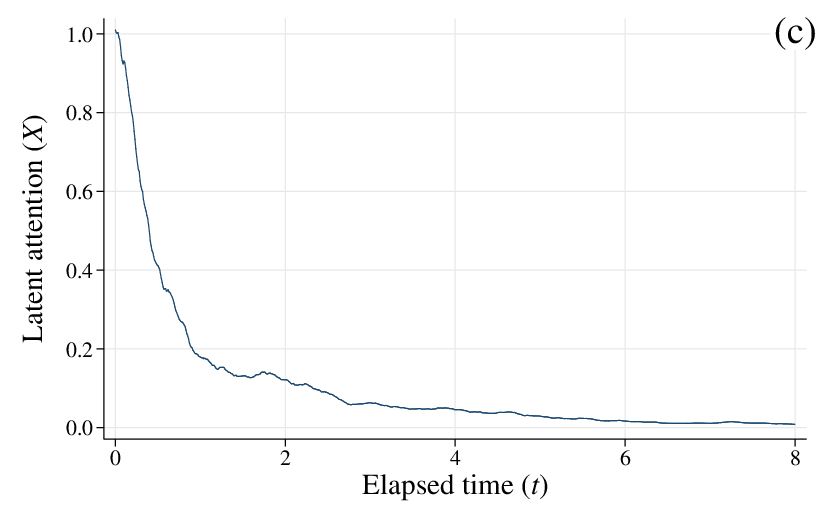}
\caption{Simulation results of the attention curve $\hat{X}(t)$ for a persistent system ($H=0.85$).  
The deterministic component was set as $\alpha(t)=1$ (constant).  
The volatility values are (a) $\beta=0.2$, (b) $\beta=0.8$ and (c) $\beta=1.5$.}
\label{fig:pers_path}
\end{minipage}
\hspace{1.5em}
\begin{minipage}[t]{0.48\textwidth}
\includegraphics[width=0.87\textwidth]{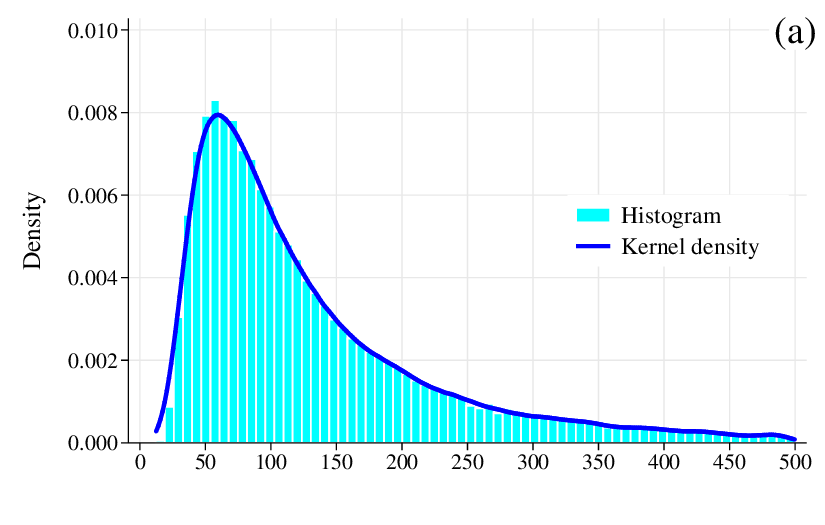}\\
\includegraphics[width=0.87\textwidth]{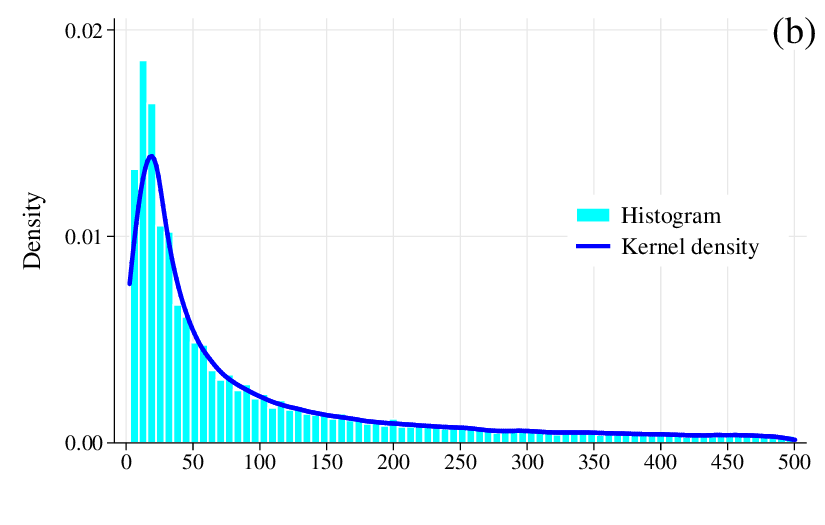}\\
\includegraphics[width=0.87\textwidth]{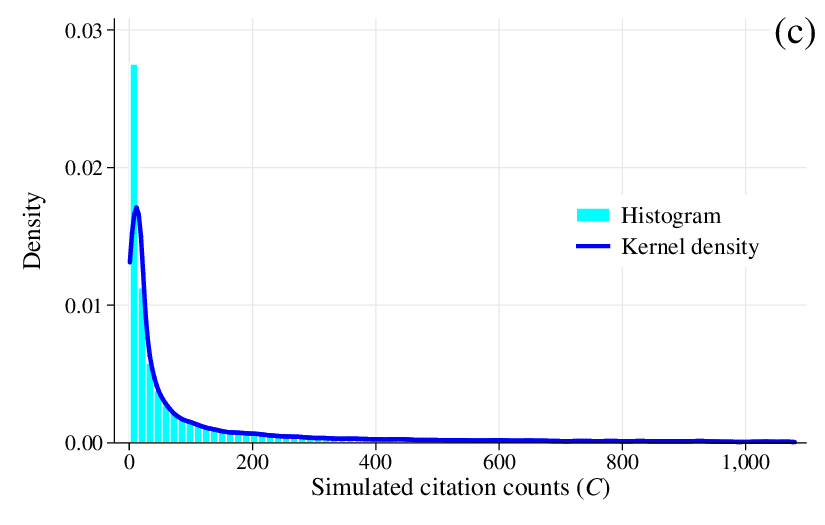}
\caption{Monte Carlo simulation results for the distribution of $\hat{C}(T)$, generated using the same parameter settings as in Fig.~\ref{fig:pers_path}, with $N_\mathrm{s}=50{,}000$ trials.  
The volatility values are (a) $\beta=0.2$, (b) $\beta=0.8$ and (c) $\beta=1.5$.}
\label{fig:pers_histo}
\end{minipage}
\end{figure}

\begin{figure}[t!]
\centering
\includegraphics[width=0.49\linewidth]{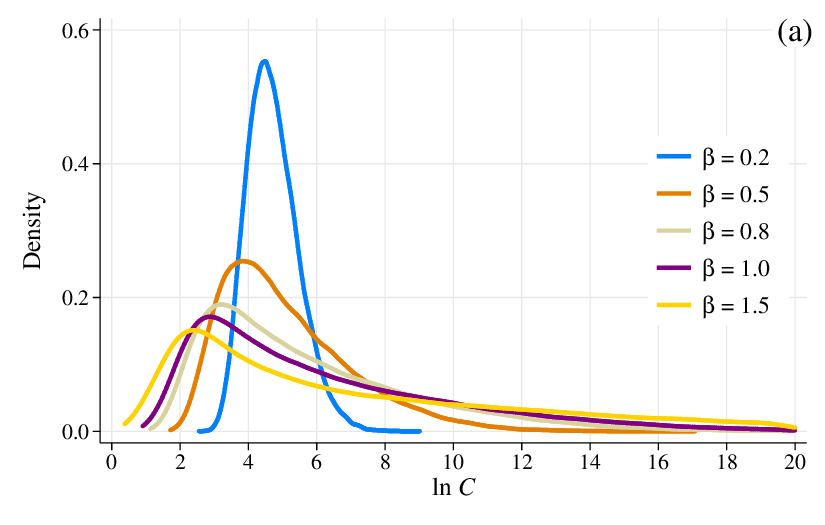}\\
\vspace{1.5em}
\includegraphics[width=0.49\linewidth]{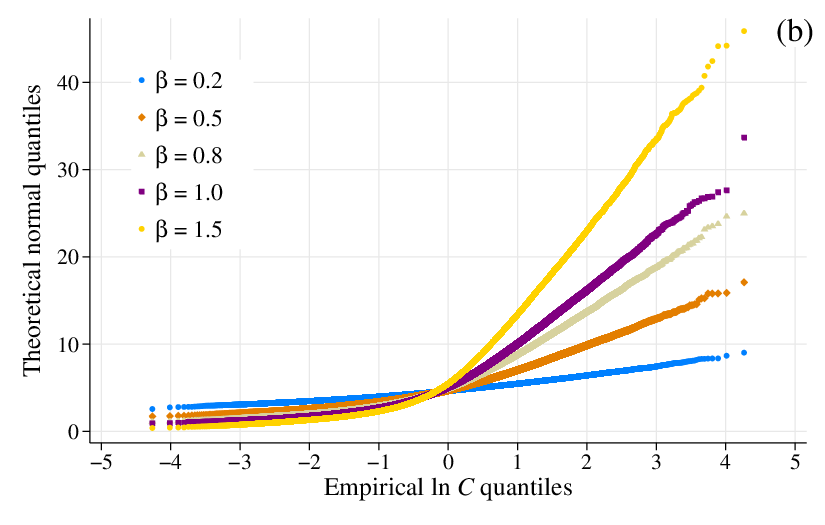}\\
\vspace{1.5em}
\includegraphics[width=0.49\linewidth]{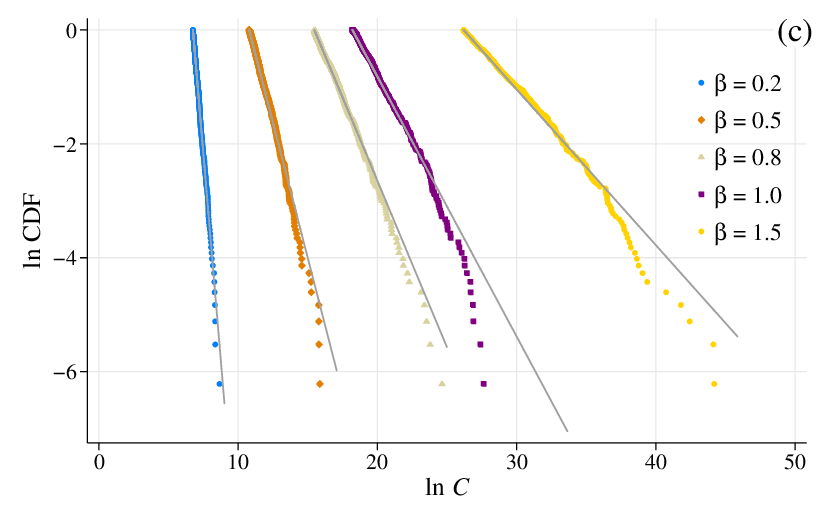}
\caption{(a) Kernel density estimation of the distribution of $\ln\hat{C}(T)$ using the same data as in Fig.~\ref{fig:pers_histo}, overlaid with different values of volatility ($\beta$) for visual clarity.
(b) Q--Q plot of the same $\ln\hat{C}(T)$ distribution against theoretical normal quantiles.  
(c) Fitting results for the upper 1\% $(N=500)$ of the $\hat{C}(T)$ distribution using a shifted power law, $P(x)\sim (x+x_{\min})^{-\gamma}$, with estimated exponents ($\hat{\gamma}$) provided in Table~\ref{tab:pers_powerlaw}.}
\label{fig:pers_QQ}
\end{figure}

\begin{figure}[t!]
\includegraphics[width=0.97\linewidth]{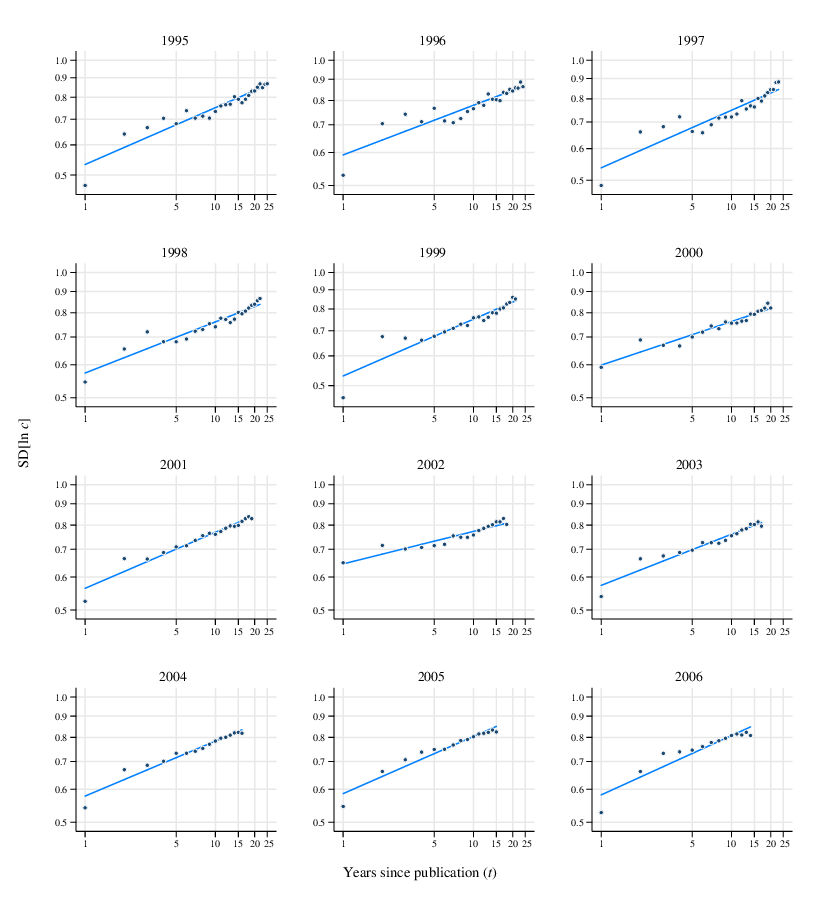}
\caption{Relationship between the standard deviation (SD) of the logarithm of citation counts acquired each year up to February 2020 and the number of years ($t$) elapsed since posting, shown for e-prints published on arXiv between 1995 and 2006.
The data used are the same as in Ref.~\cite{Okamura21}.}
\label{fig:arXiv_allyear}
\end{figure}

\begin{figure}[t!]
\centering
\includegraphics[width=0.49\linewidth]{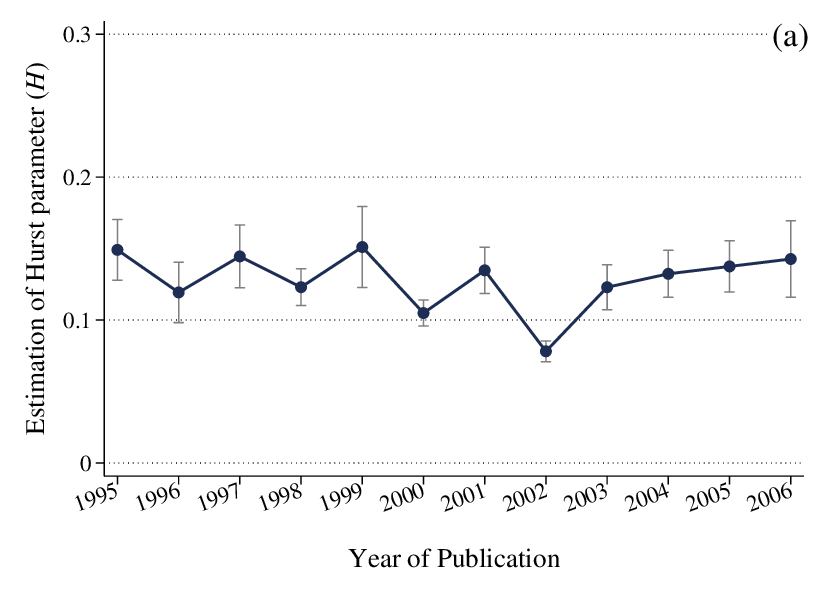}
\hspace{0.5em}
\includegraphics[width=0.49\linewidth]{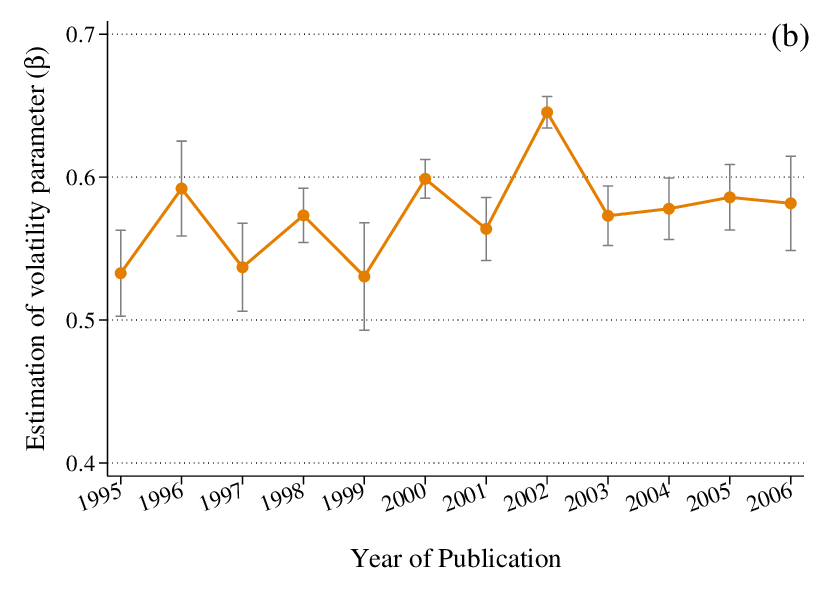}
\caption{Estimated values of (a) $\hat{H}$ and (b) $\hat{\beta}$, along with their standard errors for e-prints published on arXiv from 1995 to 2006, presented for each publication year (numerical data are provided in Table~\ref{tab:H-beta_hat}).
The data used are the same as in Ref.~\cite{Okamura21}.}
\label{fig:arXiv_allfield}
\end{figure}

\begin{figure}[t!]
\centering
\includegraphics[width=0.49\linewidth]{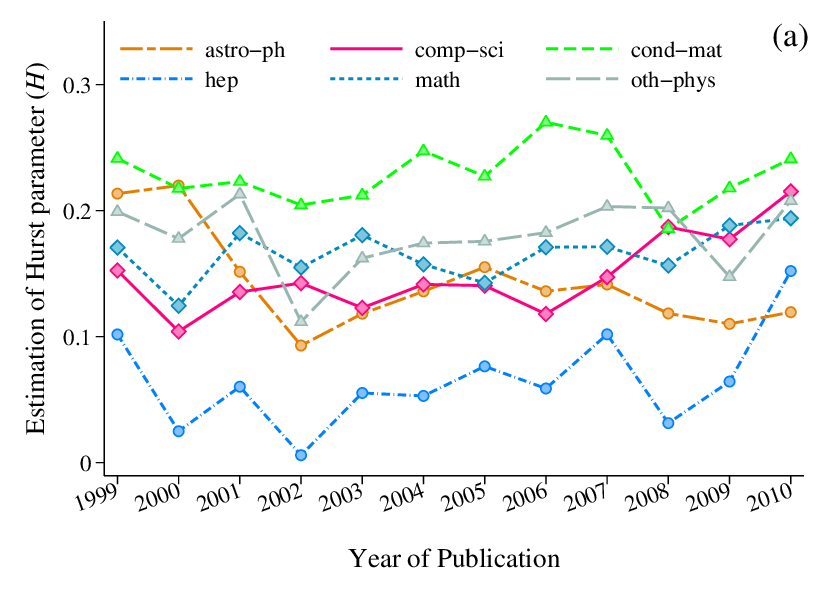}
\hspace{0.5em}
\includegraphics[width=0.49\linewidth]{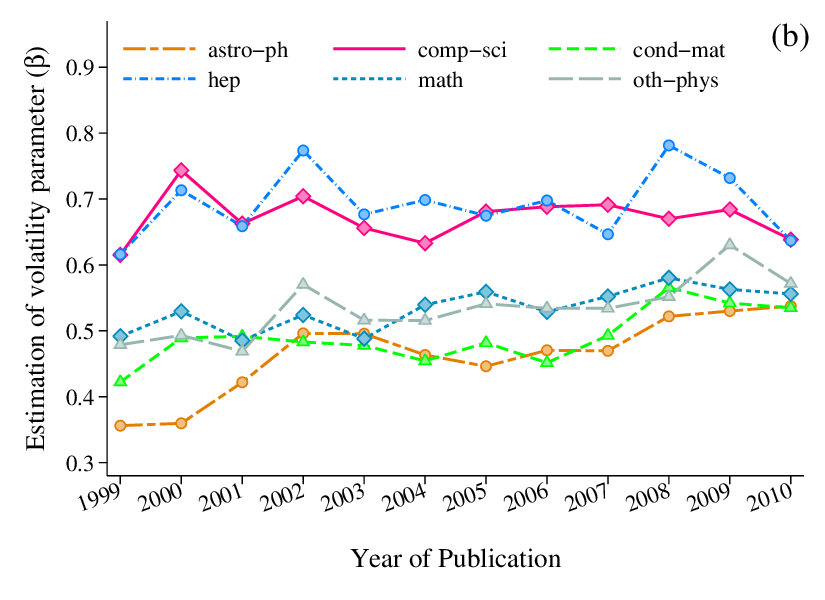}
\caption{Estimated values of (a) $\hat{H}$ and (b) $\hat{\beta}$ for e-prints published on arXiv from 1999 to 2010, shown for six different arXiv disciplines---Astrophysics (`astro-ph'), Computer Science (`comp-sci'), Condensed Matter Physics (`cond-mat'), High Energy Physics (`hep'), Mathematics (`math') and other physics (`oth-phys').
The data used are the same as in Ref.~\cite{Okamura21}.}
\label{fig:arXiv_byfield}
\end{figure}

\vspace{0.8em}
\noindent
It contains:
\vspace{-0.1em}
\begin{itemize}[leftmargin=2em]
\item Supplementary text for Secs.~\ref{sec:preliminary}--\ref{sec:discussion}\\[-1.7em]
\item Figs.~\ref{fig:pers_path}--\ref{fig:arXiv_byfield}\\[-1.7em]
\item Tables~\ref{tab:pers_powerlaw}--\ref{tab:H-beta_hat} 
\end{itemize}

\vspace{0.6em}
\noindent\dotfill
\vspace{1em}
\begin{center}
\textbf{\fontsize{11pt}{12pt}\selectfont%
Supplementary Text for Sec.~\ref{sec:preliminary}}
\end{center}
\vspace{0.5em}\noindent\textbf{$\blacktriangleright$~%
Interpretation of the $\bmk{t^{H}}$ law}\\[-1em]

The $t^{H}$ law~\eqref{law:t^H} discussed in the main text could, in principle, be reproduced within the stochastic model introduced in Ref.~\cite{Okamura22}, which extends the Black--Scholes model by incorporating a time-dependent volatility function. This can be achieved by appropriately tuning the volatility function to align with the $t^{H}$ law.  
Under this interpretation, the time dependence of volatility may be attributed to factors such as improved discoverability due to technological advancements or, conversely, reduced visibility caused by the rapid growth in publication volume. In this sense, the model provides a potential explanation for the observed phenomenon.

However, this explanation is unsatisfactory from the standpoint of theoretical universality and predictive power, as it does not fundamentally clarify the dynamical origin of the time-dependent volatility function.
For this reason, the main text adopts a more principled theoretical framework, attributing the origin of the $t^{H}$ law not to an arbitrarily imposed time-dependent volatility but to the intrinsic properties of the underlying stochastic process.
Specifically, $H$ is interpreted as the Hurst parameter of a fractional Brownian motion, and the model is formulated under the assumption of constant volatility.

Note also that the empirically estimated upper bound of the Hurst exponent for the arXiv data, as reported in Sec.~\ref{sec:analysis} of the main text, is relatively small, around $H\approx 0.13$.
When $H$ is sufficiently small, the power-law growth $t^H$ can be approximated by $1+H\ln t$.
This approximation helps explain the near-logarithmic behaviour observed in the temporal evolution of citation variance, as previously illustrated in Ref.~\cite[Appendix~C, Suppl.~Fig.~S9]{Okamura22}.

\vspace{1em}
\noindent\dotfill
\vspace{1em}
\begin{center}
\textbf{\fontsize{11pt}{12pt}\selectfont%
Supplementary Text for Sec.~\ref{sec:theory}}
\end{center}
\vspace{0.5em}\noindent\textbf{$\blacktriangleright$~%
Fractional Brownian motion via Brownian motion}\\[-1em]

There are several ways to describe a fractional Brownian motion, $\{B_{H}(t)\}_{t\in\mathbb{R}}$, with Hurst parameter $H\in(0,1)$, in terms of the standard Brownian motion, $\{W(t)\}_{t\in\mathbb{R}}$; see Ref.~\cite[Sec.~1.2]{Biagini08} and the references therein.  
Among these approaches, we introduce one that employs an operator $M_{H}$ acting on a rapidly decreasing smooth function $f$ defined on $\mathbb{R}$, explicitly given by \cite{Biagini04}
\begin{equation}\label{def:M_H}
M_{H}f(s)=
\begin{dcases*}
\,C_{H}\int_{\mathbb{R}}\abs{u}^{H-\frac{3}{2}}\big[f(s-u)-f(s)\big]\dd u & for $H\in (0,\tfrac{1}{2})$\,, \\
\,f(s) & for $H=\frac{1}{2}$\,, \\
\,C_{H}\int_{\mathbb{R}}\abs{u-s}^{H-\frac{3}{2}}f(u)\dd u & for $H\in (\tfrac{1}{2},1)$\,,
\end{dcases*}\quad
C_{H}=\frac{\sqrt{\Gamma(2H+1)\sin(\pi H)}}{2\Gamma\big(H-\frac{1}{2}\big)\cos\big[\frac{\pi}{2}\big(H-\frac{1}{2}\big)\big]}\,,
\end{equation}
where $\Gamma(\cdot)$ represents the gamma function.
Choose $f(s)$ as the indicator function $\chi_{[a,b]}(s)$, which equals $1$ if $a\leq s\leq b$, $-1$ if $b\leq s\leq a$ (except when $s=a=b$), and $0$ otherwise.  
We then integrate Eq.~\eqref{def:M_H} over the entire real line with respect to $\dd W(s)$.
Setting $a=0$ and $b=t$, we obtain
\begin{align}\label{M-VN}
\int_{\mathbb{R}}M_{H}\chi_{[0,t]}(s)\dd W(s)
=\frac{C_{H}}{\abs{H-\frac{1}{2}}}\int_{\mathbb{R}}\Big[(t-s)^{H-\frac{1}{2}}\Theta(t-s)-(-s)^{H-\frac{1}{2}}\Theta(-s)\Big]\dd W(s)\,,
\end{align}
where $\Theta(\cdot)$ denotes the Heaviside step function, defined as $\Theta(x)=1$ for $x>0$ and $\Theta(x)=0$ for $x\leq 0$.  
Equation~\eqref{M-VN} matches, up to a normalisation factor, the original expression for $B_{H}(t)$ introduced by Mandelbrot and Van Ness \cite[Eq.~(2.1)]{Mandelbrot68}.

\vspace{1.5em}\noindent\textbf{$\blacktriangleright$~%
Generalised fractional geometric Brownian motion}\\[-1em]

Consider the following generalised fractional SDE that governs the motion of a continuous random variable $X(t)$:
\begin{equation}\label{GfSDE}
dX(t)=\alpha(t)X(t)\dd t+\beta(t)X(t)\dd B_{H}(t)\,,
\end{equation}
where $\alpha(\cdot)$ and $\beta(\cdot)$ are locally bounded deterministic functions.
Assuming that $X(0)\equiv X_{0}$ is deterministic, the general solution to Eq.~\eqref{GfSDE} is given by \cite{Biagini04}
\begin{align}\label{GfSDE_sol}
X(t)&=X_{0}\exp\bigg(\int_{0}^{t}\alpha(s)\dd s+\int_{0}^{t}\beta(s)\dd B_{H}(s)
-\frac{1}{2}\int_{\mathbb{R}}\big[M_{H}(\beta\circ\chi_{[0,t]})(s)\big]^{2}\dd s\bigg)\,,
\end{align}
where the operator $M_{H}(\cdot)$ and the indicator function $\chi_{[a,b]}(\cdot)$ are as defined earlier.
In the main text, we considered the special case where $\beta(s)=\beta$ is constant, and under that condition, the solution~\eqref{GfSDE_sol} reduces to Eqs.~\eqref{fSDE_sol} and \eqref{A_t}.

\vspace{1em}
\noindent\dotfill
\vspace{1em}
\begin{center}
\textbf{\fontsize{11pt}{12pt}\selectfont%
Supplementary Text for Sec.~\ref{sec:simulation}}
\end{center}
\vspace{0.5em}\noindent\textbf{$\blacktriangleright$~%
From attention to citation: a pragramatic approach}\\[-1em]

In the main text, when simulating $C(T)$ as expressed in Eq.~\eqref{C<->X} based on our theoretical model, we introduced the approximation $\hat{C}(T)$ as given in Eq.~\eqref{def:C_tilde}.
This approximation always yields a natural number for $T>0$, even in cases where $X_{i}\dD t$ remains negligibly small throughout $i=1,\,\dots,\,n$, such that its summation over the entire period does not even reach 1.
Nonetheless, Eq.~\eqref{def:C_tilde} ensures that $\hat{C}=1$ even in such situations.
In this respect, the effect is analogous to the `$\text{citation}+1$' prescription often used in the literature \cite{Seglen92,Golosovsky12b,Thelwall16b,Okamura22}, which facilitates the computation of the logarithm of citation counts even in cases with zero citations.
We assume that the bias introduced by employing the ceiling function here is negligible and does not affect our main conclusions.

\vspace{1.5em}\noindent\textbf{$\blacktriangleright$~%
Simulation of persistent cases}\\[-1em]

It is discussed in the main text that the conditions for a good fit by a log-normal distribution and by a power law contradict each other in several respects, particularly regarding whether $H$ is greater or less than $\tfrac{1}{2}$, the background trend determined by $u(t)$, which is related to the drift function $\alpha(t)$ by Eq.~\eqref{alpha<->u}, and the magnitude of volatility.
To gain an intuitive grasp of this point, we conducted an additional simulation under a hypothetical scenario where the system exhibits a persistent memory effect and the function $u(t)$ does not decay.
The results are summarised below.

Figure~\ref{fig:pers_path} shows simulated paths of $X(t)$ under $H=0.85$ and $\alpha(t)=1$.  
In contrast to the antipersistent case (Fig.~\ref{fig:antipers_path}), the trajectories here are notably smoother, with fewer pronounced fluctuations.  
In Fig.~\ref{fig:pers_path}(c), $\{X_{i}\}_{i=1}^{n}$ follows a nearly steady trend and eventually declines to zero at time $T$.  
However, this outcome is not inevitable.  
Due to the persistent nature of the process, $\{X_{i}\}_{i=1}^{n}$ often continues along a certain trajectory, yet it could also move in the opposite direction and increase indefinitely.  
Such paths contribute to large values of $\hat{C}(T)$.

Figure~\ref{fig:pers_histo} shows the results of Monte Carlo simulations with $N_\mathrm{s}=50{,}000$, conducted under the same conditions.  
Additionally, Fig.~\ref{fig:pers_QQ}(a) presents the kernel density estimates of the distribution for $\beta=0.2$, $0.5$, $0.8$, $1.0$ and $1.5$.
Compared with the antipersistent case (Fig.~\ref{fig:antipers_histo}), the distribution of $C(T)$ now has a substantially fatter tail, a tendency that grows more pronounced with increasing $\beta$.

Moreover, the Q--Q plot in Fig.~\ref{fig:pers_QQ}(b) deviates noticeably from a straight line, indicating that the distribution of $\hat{C}(T)$ no longer closely follows a log-normal form.  
This does not necessarily mean that a power law is the best fit.  
However, as shown in Fig.~\ref{fig:pers_QQ}(c) and Table~\ref{tab:pers_powerlaw}, where the upper 1\% of $\hat{C}(T)$ is fitted with a shifted power law, the persistent scenario exhibits a stronger tendency to be well approximated by a shifted power law than the antipersistent case.  
Still, at extremely high citation counts, the data also deviate from the shifted power law.

\vspace{1em}
\noindent\dotfill
\vspace{1em}
\begin{center}
\textbf{\fontsize{11pt}{12pt}\selectfont%
Supplementary Text for Sec.~\ref{sec:analysis}}
\end{center}
\vspace{0.5em}\noindent\textbf{$\blacktriangleright$~%
The arXiv data}\\[-1em]

The data used in Fig.~\ref{fig:t^H} and the empirical data analysis in Sec.~\ref{sec:analysis} of the main text are the same as those employed in Ref.~\cite{Okamura22}, with the raw data compiled in Ref.~\cite{Okamura21}.  
For further details on the data, please refer to these references.

\vspace{1.5em}\noindent\textbf{$\blacktriangleright$~%
Prospective vs.\ retrospective approach}\\[-1em]

To verify the $t^{H}$ law, it is necessary to examine data on the standard deviation (SD) of the logarithm of $c(t;\Delta t)\equiv\Delta C(t)/\Delta t$.  
In conceptualising or formalising how this measurement is carried out in practice, two approaches can be considered.

The first approach, referred to as the \emph{prospective} approach, fixes the publication year of the analysed set of publications while progressively shifting forward the years in which citations are counted.  
In other words, for a set of publications from the same year, it tracks how the SD of the logarithm of citation counts per unit time evolves over time.
More concretely, defining $t_{k}=k\Delta t$ $(k=0,\,\dots,\,m)$, and denoting by $\mathcal{L}_{k}$ the set of publications published during $(t_{k-1},t_{k}]$, and by $c_{w,k}$ the number of citations a publication $w$ receives in $(t_{k-1},t_{k}]$, the SD is calculated for the dataset $\{\ln c_{w,k}\,|\,w\in \mathcal{L}_{1}\}$ for each $k=1,\,\dots,\,m$.  
Under this method, the size of the publication set under study remains $\abs{\mathcal{L}_{1}}$.

The second approach, referred to as the \emph{retrospective} approach, instead fixes the year in which citations are counted and defines the analysed set of publications by tracing back their publication years.
Using the same notation as above, the SD is computed for the dataset $\{\ln c_{w,m}\,|\,w\in \mathcal{L}_{m-k}\}$ for each $k=0,\,\dots,\,m-1$.  
In this case, the size of the analysed set of publications ($\abs{\mathcal{L}_{m-k}}$) typically increases as $k$ decreases, owing to the growth in the number of publications over time.

In the main text, the prospective approach is used (Sec.~\ref{sec:law}).  
However, using the retrospective approach yields qualitatively similar results, again confirming the $t^{H}$ law trend.  
Indeed, when the $t^{H}$ law was first observed, the retrospective approach was applied \cite[Appendix C, Suppl.~Fig.~S9]{Okamura22}.

\vspace{1.5em}\noindent\textbf{$\blacktriangleright$~%
Estimation results by discipline}\\[-1em]

In the main text, the estimation of the Hurst parameter $H$ and volatility $\beta$ was carried out on the entire aggregate set of e-prints across six arXiv disciplines---Astrophysics (`astro-ph'), Computer Science (`comp-sci'), Condensed Matter Physics (`cond-mat'), High Energy Physics (`hep'), Mathematics (`math') and other physics (`oth-phys')---the same classification as in Refs.~\cite{Okamura22,Okamura21}.  
Figure~\ref{fig:arXiv_byfield} shows the results of the same analysis, now broken down by discipline.
From Fig.~\ref{fig:arXiv_byfield}(a), it is clear that the memory effect is antipersistent in all fields, but a consistent historical trend indicates that High Energy Physics has the strongest antipersistent tendency, whereas Condensed Matter Physics has the weakest.  
Additionally, Fig.~\ref{fig:arXiv_byfield}(b) reveals that volatility is especially high in High Energy Physics and Computer Science.

These observations likely reflect each discipline's distinct characteristics, such as scientific discovery patterns, citation behaviours in research communities, and publishing norms---factors also discussed in the main text.  
Combined with earlier analyses of discipline-specific spatiotemporal aspects of citation evolution and with measures such as the `internal obsolescence rate' and `retention rate' reported in Ref.~\cite{Okamura22}, these results provide useful insights not only from scientometric or bibliometric standpoints, but also within the broader `science of science' framework \cite{Fortunato18}.

\newpage
\vspace{1.5em}\noindent\textbf{$\blacktriangleright$~%
Limitations and validity}\\[-1em]

Typically, the Hurst parameter for fractional Brownian motion is estimated via methods such as the rescaled range analysis (R/S) method \cite{Hurst51,Mandelbrot72} or subsequent methodological refinements.
This approach involves calculating the range of partial sums of deviations for time series segments from their means, then normalising by the corresponding SD.
However, it is not suitable in te current study.  
The reason is that the presence of an unknown, time-dependent deterministic nonlinear term $\alpha(t)$ disrupts self-similar scaling properties, thereby biasing R/S-based estimates.

Moreover, the arXiv dataset used in this study \cite{Okamura21} records citation counts on a yearly basis, yielding at most a few dozen data points per e-print.  
Methods like R/S analysis \cite{Hurst51,Mandelbrot72}, detrended fluctuation analysis (DFA) \cite{Peng94,*Hu01,*Chen02}, or alternative estimation techniques commonly employed in financial engineering \cite{Biagini08,Pipiras17} usually require much longer time series.
Despite recent advances in bibliometric data, collecting extended citation histories at finer temporal resolutions (e.g.\ monthly or daily) remains practically challenging.
Where monthly-level citation records exist, they often involve data imputation: for instance, if the exact publication month is missing, the record may default to January.  
Similarly, daily-level data can default to the first of the month if only the month is known.  
Hence, careful preprocessing is needed to handle missing values or exclude certain entries, making the construction of precise time series difficult at monthly or daily resolutions.

Given these constraints, this study employed a heuristic estimation of the Hurst parameter based on the newly discovered $t^{H}$ law.
Although the resulting estimates are approximate, their qualitative features---such as the evidence of antipersistent behaviour---are statistically robust within the limits of the applied tests.
It should nevertheless be emphasised that these results are derived from a simplified estimation procedure and from data with a limited number of points, and must therefore be interpreted with due regard to the associated biases and limitations.

Regarding future research directions, we anticipate that as bibliometric data collection expands in both scope and accuracy across various institutions and platforms, more reliable long-term citation and attention-based datasets will become available.  
This advancement will allow for more precise estimation of parameters such as $H$ and $\beta$, further enriching our understanding of citation dynamics.

\vspace{1em}
\noindent\dotfill
\vspace{1em}
\begin{center}
\textbf{\fontsize{11pt}{12pt}\selectfont%
Supplementary Text for Sec.~\ref{sec:discussion}}
\end{center}
\vspace{0.5em}\noindent\textbf{$\blacktriangleright$~%
`Unlimited attention capability': a hypothetical scenario}\\[-1em]

With anticipated advancements in artificial intelligence technologies and the continual evolution of digital platforms, it is worth considering a hypothetical scenario that, while speculative, could emerge in the future.
Empirically, the average attention history curve we have observed follows a jump-and-decay pattern: newly published research can initially attract significant attention, but citations gradually diminish over time.
Various factors contribute to this decline, including constraints on reference lists (e.g., space limitations or submission guidelines), the cognitive limits of researchers and authors, and the phenomenon of `obliteration by incorporation' \cite{Garfield75}, where widely accepted findings become so foundational that they are seldom explicitly cited.

Now consider a hypothetical world in which all publishing constraints---whether resource-based, technological, or conventional---are entirely removed.
In this setting, researchers would have optimal access to literature search, knowledge acquisition, and reference management.
Citing older publications would be actively encouraged, with no practical limit on the number of references included.
Under such conditions, the typical citation history curve, when aggregated across numerous publications, might no longer follow a jump-and-decay pattern but instead develop a heavy-tailed form over time or even evolve into a monotonically increasing function.
In this scenario, the distribution of the time-integrated attention measure would likely be better represented by a power law or another even heavier-tailed distribution, rather than a log-normal distribution.
Alternatively, it is possible that such an attention-dynamics paradigm already exists in certain domains, even if it has not yet been widely recognised.

\clearpage
\begin{center}
\textbf{\fontsize{11pt}{12pt}\selectfont%
Supplementary Figures}
\end{center}
\vspace{1em}


\clearpage

\clearpage

\clearpage
\begin{center}
\textbf{\fontsize{11pt}{12pt}\selectfont%
Supplementary Tables}
\end{center}
\vspace{1em}

\begin{table}[h!]
\caption{Fitting results for the upper 1\% ($N=500$) of the $\hat{C}(T)$ distribution for different values of $\beta$, based on the same data as in Fig.~\ref{fig:pers_histo}, using a shifted power law: $P(x)\sim (x+x_{\min})^{-\gamma}$.  
All cases satisfy $p<0.001$.}
\label{tab:pers_powerlaw}
\setlength{\tabcolsep}{10pt}
\vspace{1.0em}
\begin{tabular}{cc}
\midrule\addlinespace[-0.1ex]
\midrule\\[-1.65em]
$\beta$ & $\hat{\gamma}$ \\[-0.2em]
\midrule
0.2 & ~\,$2.9 \pm 3.0\times 10^{-2}$ \\
0.5 & ~\,$1.0 \pm 1.2\times 10^{-2}$ \\
0.8 & $0.62 \pm 4.8\times 10^{-3}$ \\
1.0 & $0.50 \pm 8.6\times 10^{-3}$ \\
1.5 & $0.30 \pm 3.8\times 10^{-3}$ \\
\midrule\addlinespace[-0.1ex]
\midrule
\end{tabular}
\end{table}
\begin{table}[h!]
\caption{Summary of the estimated Hurst parameter ($\hat{H}$) and volatility parameter ($\hat{\beta}$) for each publication year, based on the same data as in Fig.~\ref{fig:arXiv_allfield}.
All cases satisfy $p<0.001$.}
\label{tab:H-beta_hat}
\setlength{\tabcolsep}{10pt}
\vspace{1.0em}
\begin{tabular}{cccc}
\midrule\addlinespace[-0.1ex]
\midrule\\[-1.45em]
Year & \#\,e-print & $\hat{H}$ & $\hat{\beta}$ \\[-0.2em]
\midrule
1995 & 13{,}618 & $0.15 \pm 0.021$ & $0.53 \pm 0.030$ \\
1996 & 16{,}869 & $0.12 \pm 0.021$ & $0.59 \pm 0.033$ \\
1997 & 22{,}306 & $0.14 \pm 0.022$ & $0.54 \pm 0.031$ \\
1998 & 28{,}016 & $0.12 \pm 0.013$ & $0.57 \pm 0.019$ \\
1999 & 32{,}091 & $0.15 \pm 0.028$ & $0.53 \pm 0.038$ \\
2000 & 36{,}173 & $0.10 \pm 0.009$ & $0.60 \pm 0.014$ \\
2001 & 39{,}106 & $0.13 \pm 0.016$ & $0.56 \pm 0.022$ \\
2002 & 42{,}642 & $0.08 \pm 0.007$ & $0.65 \pm 0.011$ \\
2003 & 46{,}589 & $0.12 \pm 0.016$ & $0.57 \pm 0.021$ \\
2004 & 52{,}534 & $0.13 \pm 0.016$ & $0.58 \pm 0.022$ \\
2005 & 56{,}801 & $0.14 \pm 0.018$ & $0.59 \pm 0.023$ \\
2006 & 61{,}177 & $0.14 \pm 0.027$ & $0.58 \pm 0.033$ \\
\midrule\addlinespace[-0.1ex]
\midrule
\end{tabular}
\end{table}


\begin{thebibliography}{87}%
\makeatletter
\providecommand \@ifxundefined [1]{%
 \@ifx{#1\undefined}
}%
\providecommand \@ifnum [1]{%
 \ifnum #1\expandafter \@firstoftwo
 \else \expandafter \@secondoftwo
 \fi
}%
\providecommand \@ifx [1]{%
 \ifx #1\expandafter \@firstoftwo
 \else \expandafter \@secondoftwo
 \fi
}%
\providecommand \natexlab [1]{#1}%
\providecommand \enquote  [1]{``#1''}%
\providecommand \bibnamefont  [1]{#1}%
\providecommand \bibfnamefont [1]{#1}%
\providecommand \citenamefont [1]{#1}%
\providecommand \href@noop [0]{\@secondoftwo}%
\providecommand \href [0]{\begingroup \@sanitize@url \@href}%
\providecommand \@href[1]{\@@startlink{#1}\@@href}%
\providecommand \@@href[1]{\endgroup#1\@@endlink}%
\providecommand \@sanitize@url [0]{\catcode `\\12\catcode `\$12\catcode
  `\&12\catcode `\#12\catcode `\^12\catcode `\_12\catcode `\%12\relax}%
\providecommand \@@startlink[1]{}%
\providecommand \@@endlink[0]{}%
\providecommand \url  [0]{\begingroup\@sanitize@url \@url }%
\providecommand \@url [1]{\endgroup\@href {#1}{\urlprefix }}%
\providecommand \urlprefix  [0]{URL }%
\providecommand \Eprint [0]{\href }%
\providecommand \doibase [0]{https://doi.org/}%
\providecommand \selectlanguage [0]{\@gobble}%
\providecommand \bibinfo  [0]{\@secondoftwo}%
\providecommand \bibfield  [0]{\@secondoftwo}%
\providecommand \translation [1]{[#1]}%
\providecommand \BibitemOpen [0]{}%
\providecommand \bibitemStop [0]{}%
\providecommand \bibitemNoStop [0]{.\EOS\space}%
\providecommand \EOS [0]{\spacefactor3000\relax}%
\providecommand \BibitemShut  [1]{\csname bibitem#1\endcsname}%
\let\auto@bib@innerbib\@empty
\bibitem [{\citenamefont {Einstein}(1905)}]{Einstein05}%
  \BibitemOpen
  \bibfield  {author} {\bibinfo {author} {\bibfnamefont {A.}~\bibnamefont
  {Einstein}},\ }\href {https://doi.org/10.1002/andp.19053220806} {\bibfield
  {journal} {\bibinfo  {journal} {Ann.~Phys.}\ }\textbf {\bibinfo {volume}
  {17}},\ \bibinfo {pages} {549} (\bibinfo {year} {1905})}\BibitemShut
  {NoStop}%
\bibitem [{\citenamefont {Mandelbrot}(1982)}]{Mandelbrot82}%
  \BibitemOpen
  \bibfield  {author} {\bibinfo {author} {\bibfnamefont {B.~B.}\ \bibnamefont
  {Mandelbrot}},\ }\href@noop {} {\emph {\bibinfo {title} {The Fractal Geometry
  of Nature}}}\ (\bibinfo  {publisher} {W.~H.~Freeman \& Co.},\ \bibinfo
  {address} {New York},\ \bibinfo {year}
  {1982})\BibitemShut {NoStop}%
\bibitem [{\citenamefont {Falconer}(2014)}]{Falconer14}%
  \BibitemOpen
  \bibfield  {author} {\bibinfo {author} {\bibfnamefont {K.}~\bibnamefont
  {Falconer}},\ }\href@noop {} {\emph {\bibinfo {title} {Fractal Geometry:
  Mathematical Foundations and Applications}}},\ \bibinfo {edition} {3rd}\ ed.\
  (\bibinfo  {publisher} {John Wiley \& Sons, Ltd},\ \bibinfo
  {address} {New York},\ \bibinfo {year}
  {2014})\BibitemShut {NoStop}%
\bibitem [{\citenamefont {Turing}(1952)}]{Turing52}%
  \BibitemOpen
  \bibfield  {author} {\bibinfo {author} {\bibfnamefont {A.~M.}\ \bibnamefont
  {Turing}},\ }\href {https://doi.org/10.1098/rstb.1952.0012} {\bibfield
  {journal} {\bibinfo  {journal} {Philos.~Trans.~R.~Soc.~B}\ }\textbf {\bibinfo
  {volume} {237}},\ \bibinfo {pages} {37} (\bibinfo {year} {1952})}\BibitemShut
  {NoStop}%
\bibitem [{\citenamefont {Keller}\ and\ \citenamefont
  {Segel}(1971)}]{Keller71}%
  \BibitemOpen
  \bibfield  {author} {\bibinfo {author} {\bibfnamefont {E.~F.}\ \bibnamefont
  {Keller}}\ and\ \bibinfo {author} {\bibfnamefont {L.~A.}\ \bibnamefont
  {Segel}},\ }\href {https://doi.org/10.1016/0022-5193(71)90050-6} {\bibfield
  {journal} {\bibinfo  {journal} {J.~Theor.~Biol.}\ }\textbf {\bibinfo {volume}
  {30}},\ \bibinfo {pages} {225} (\bibinfo {year} {1971})}\BibitemShut
  {NoStop}%
\bibitem [{\citenamefont {Watts}\ and\ \citenamefont
  {Strogatz}(1998)}]{Watts98}%
  \BibitemOpen
  \bibfield  {author} {\bibinfo {author} {\bibfnamefont {D.~J.}\ \bibnamefont
  {Watts}}\ and\ \bibinfo {author} {\bibfnamefont {S.~H.}\ \bibnamefont
  {Strogatz}},\ }\href {https://doi.org/10.1038/30918} {\bibfield  {journal}
  {\bibinfo  {journal} {Nature}\ }\textbf {\bibinfo {volume} {393}},\ \bibinfo
  {pages} {440} (\bibinfo {year} {1998})}\BibitemShut {NoStop}%
\bibitem [{\citenamefont {Albert}\ and\ \citenamefont
  {Barab\'{a}si}(2002)}]{Albert02}%
  \BibitemOpen
  \bibfield  {author} {\bibinfo {author} {\bibfnamefont {R.}~\bibnamefont
  {Albert}}\ and\ \bibinfo {author} {\bibfnamefont {A.-L.}\ \bibnamefont
  {Barab\'{a}si}},\ }\href {https://doi.org/10.1103/RevModPhys.74.47}
  {\bibfield  {journal} {\bibinfo  {journal} {Rev.~Mod.~Phys.}\ }\textbf
  {\bibinfo {volume} {74}},\ \bibinfo {pages} {47} (\bibinfo {year}
  {2002})}\BibitemShut {NoStop}%
\bibitem [{\citenamefont {Price}(1965)}]{Price65}%
  \BibitemOpen
  \bibfield  {author} {\bibinfo {author} {\bibfnamefont {D.~J. d.~S.}\
  \bibnamefont {Price}},\ }\href {https://doi.org/10.1126/science.149.3683.510}
  {\bibfield  {journal} {\bibinfo  {journal} {Science}\ }\textbf {\bibinfo
  {volume} {149}},\ \bibinfo {pages} {510} (\bibinfo {year}
  {1965})}\BibitemShut {NoStop}%
\bibitem [{\citenamefont {Price}(1976)}]{Price76}%
  \BibitemOpen
  \bibfield  {author} {\bibinfo {author} {\bibfnamefont {D.~J. d.~S.}\
  \bibnamefont {Price}},\ }\href {https://doi.org/10.1002/asi.4630270505}
  {\bibfield  {journal} {\bibinfo  {journal} {J.~Am.~Soc.~Inf.~Sci.}\ }\textbf
  {\bibinfo {volume} {27}},\ \bibinfo {pages} {292} (\bibinfo {year}
  {1976})}\BibitemShut {NoStop}%
\bibitem [{\citenamefont {Redner}(1998)}]{Redner98}%
  \BibitemOpen
  \bibfield  {author} {\bibinfo {author} {\bibfnamefont {S.}~\bibnamefont
  {Redner}},\ }\href {https://doi.org/10.1007/s100510050359} {\bibfield
  {journal} {\bibinfo  {journal} {Eur.~Phys.~J.~B}\ }\textbf {\bibinfo {volume}
  {4}},\ \bibinfo {pages} {131} (\bibinfo {year} {1998})}\BibitemShut {NoStop}%
\bibitem [{\citenamefont {Hajra}\ and\ \citenamefont {Sen}(2006)}]{Hajra06}%
  \BibitemOpen
  \bibfield  {author} {\bibinfo {author} {\bibfnamefont {K.~B.}\ \bibnamefont
  {Hajra}}\ and\ \bibinfo {author} {\bibfnamefont {P.}~\bibnamefont {Sen}},\
  }\href {https://doi.org/10.1016/j.physa.2005.12.044} {\bibfield  {journal}
  {\bibinfo  {journal} {Physica A}\ }\textbf {\bibinfo {volume} {368}},\
  \bibinfo {pages} {575} (\bibinfo {year} {2006})}\BibitemShut {NoStop}%
\bibitem [{\citenamefont {Lehmann}\ \emph {et~al.}(2003)\citenamefont
  {Lehmann}, \citenamefont {Lautrup},\ and\ \citenamefont
  {Jackson}}]{Lehmann03}%
  \BibitemOpen
  \bibfield  {author} {\bibinfo {author} {\bibfnamefont {S.}~\bibnamefont
  {Lehmann}}, \bibinfo {author} {\bibfnamefont {B.}~\bibnamefont {Lautrup}},\
  and\ \bibinfo {author} {\bibfnamefont {A.~D.}\ \bibnamefont {Jackson}},\
  }\href {https://doi.org/10.1103/PhysRevE.68.026113} {\bibfield  {journal}
  {\bibinfo  {journal} {Phys.~Rev.~E}\ }\textbf {\bibinfo {volume} {68}},\
  \bibinfo {pages} {026113} (\bibinfo {year} {2003})}\BibitemShut {NoStop}%
\bibitem [{\citenamefont {Clauset}\ \emph {et~al.}(2009)\citenamefont
  {Clauset}, \citenamefont {Shalizi},\ and\ \citenamefont
  {Newman}}]{Clauset09}%
  \BibitemOpen
  \bibfield  {author} {\bibinfo {author} {\bibfnamefont {A.}~\bibnamefont
  {Clauset}}, \bibinfo {author} {\bibfnamefont {C.~R.}\ \bibnamefont
  {Shalizi}},\ and\ \bibinfo {author} {\bibfnamefont {M.~E.~J.}\ \bibnamefont
  {Newman}},\ }\href {https://doi.org/10.1137/070710111} {\bibfield  {journal}
  {\bibinfo  {journal} {SIAM Rev.}\ }\textbf {\bibinfo {volume} {51}},\
  \bibinfo {pages} {661} (\bibinfo {year} {2009})}\BibitemShut {NoStop}%
\bibitem [{\citenamefont {Peterson}\ \emph {et~al.}(2010)\citenamefont
  {Peterson}, \citenamefont {Press\'e},\ and\ \citenamefont
  {Dill}}]{Peterson10}%
  \BibitemOpen
  \bibfield  {author} {\bibinfo {author} {\bibfnamefont {G.~J.}\ \bibnamefont
  {Peterson}}, \bibinfo {author} {\bibfnamefont {S.}~\bibnamefont {Press\'e}},\
  and\ \bibinfo {author} {\bibfnamefont {K.~A.}\ \bibnamefont {Dill}},\ }\href
  {https://doi.org/10.1073/pnas.1010757107} {\bibfield  {journal} {\bibinfo
  {journal} {Proc.~Natl.~Acad.~Sci.~USA}\ }\textbf {\bibinfo {volume} {107}},\
  \bibinfo {pages} {16023} (\bibinfo {year} {2010})}\BibitemShut {NoStop}%
\bibitem [{\citenamefont {Brzezinski}(2015)}]{Brzezinski15}%
  \BibitemOpen
  \bibfield  {author} {\bibinfo {author} {\bibfnamefont {M.}~\bibnamefont
  {Brzezinski}},\ }\href {https://doi.org/10.1007/s11192-014-1524-z} {\bibfield
   {journal} {\bibinfo  {journal} {Scientometrics}\ }\textbf {\bibinfo {volume}
  {103}},\ \bibinfo {pages} {213} (\bibinfo {year} {2015})}\BibitemShut
  {NoStop}%
\bibitem [{\citenamefont {Eom}\ and\ \citenamefont {Fortunato}(2011)}]{Eom11}%
  \BibitemOpen
  \bibfield  {author} {\bibinfo {author} {\bibfnamefont {Y.-H.}\ \bibnamefont
  {Eom}}\ and\ \bibinfo {author} {\bibfnamefont {S.}~\bibnamefont
  {Fortunato}},\ }\href {https://doi.org/10.1371/journal.pone.0024926}
  {\bibfield  {journal} {\bibinfo  {journal} {PLoS ONE}\ }\textbf {\bibinfo
  {volume} {6}},\ \bibinfo {pages} {e24926} (\bibinfo {year}
  {2011})}\BibitemShut {NoStop}%
\bibitem [{\citenamefont {Redner}(2005)}]{Redner05}%
  \BibitemOpen
  \bibfield  {author} {\bibinfo {author} {\bibfnamefont {S.}~\bibnamefont
  {Redner}},\ }\href {https://doi.org/10.1063/1.1996475} {\bibfield  {journal}
  {\bibinfo  {journal} {Phys.~Today}\ }\textbf {\bibinfo {volume} {58}},\
  \bibinfo {pages} {49} (\bibinfo {year} {2005})}\BibitemShut {NoStop}%
\bibitem [{\citenamefont {Radicchi}\ \emph {et~al.}(2008)\citenamefont
  {Radicchi}, \citenamefont {Fortunato},\ and\ \citenamefont
  {Castellano}}]{Radicchi08}%
  \BibitemOpen
  \bibfield  {author} {\bibinfo {author} {\bibfnamefont {F.}~\bibnamefont
  {Radicchi}}, \bibinfo {author} {\bibfnamefont {S.}~\bibnamefont
  {Fortunato}},\ and\ \bibinfo {author} {\bibfnamefont {C.}~\bibnamefont
  {Castellano}},\ }\href {https://doi.org/10.1073/pnas.0806977105} {\bibfield
  {journal} {\bibinfo  {journal} {Proc.~Natl.~Acad.~Sci.~USA}\ }\textbf
  {\bibinfo {volume} {105}},\ \bibinfo {pages} {17268} (\bibinfo {year}
  {2008})}\BibitemShut {NoStop}%
\bibitem [{\citenamefont {Evans}\ \emph {et~al.}(2012)\citenamefont {Evans},
  \citenamefont {Hopkins},\ and\ \citenamefont {Kaube}}]{Evans12}%
  \BibitemOpen
  \bibfield  {author} {\bibinfo {author} {\bibfnamefont {T.~S.}\ \bibnamefont
  {Evans}}, \bibinfo {author} {\bibfnamefont {N.}~\bibnamefont {Hopkins}},\
  and\ \bibinfo {author} {\bibfnamefont {B.~S.}\ \bibnamefont {Kaube}},\ }\href
  {https://doi.org/10.1007/s11192-012-0694-9} {\bibfield  {journal} {\bibinfo
  {journal} {Scientometrics}\ }\textbf {\bibinfo {volume} {93}},\ \bibinfo
  {pages} {473} (\bibinfo {year} {2012})}\BibitemShut {NoStop}%
\bibitem [{\citenamefont {Chatterjee}\ \emph {et~al.}(2016)\citenamefont
  {Chatterjee}, \citenamefont {Ghosh},\ and\ \citenamefont
  {Chakrabarti}}]{Chatterjee16}%
  \BibitemOpen
  \bibfield  {author} {\bibinfo {author} {\bibfnamefont {A.}~\bibnamefont
  {Chatterjee}}, \bibinfo {author} {\bibfnamefont {A.}~\bibnamefont {Ghosh}},\
  and\ \bibinfo {author} {\bibfnamefont {B.~K.}\ \bibnamefont {Chakrabarti}},\
  }\href {https://doi.org/10.1371/journal.pone.0148863} {\bibfield  {journal}
  {\bibinfo  {journal} {PLoS ONE}\ }\textbf {\bibinfo {volume} {11}},\ \bibinfo
  {pages} {e0148863} (\bibinfo {year} {2016})}\BibitemShut {NoStop}%
\bibitem [{\citenamefont {D'Angelo}\ and\ \citenamefont {{Di
  Russo}}(2019)}]{DAngelo19}%
  \BibitemOpen
  \bibfield  {author} {\bibinfo {author} {\bibfnamefont {C.~A.}\ \bibnamefont
  {D'Angelo}}\ and\ \bibinfo {author} {\bibfnamefont {S.}~\bibnamefont {{Di
  Russo}}},\ }\href {https://doi.org/10.1016/j.joi.2019.03.011} {\bibfield
  {journal} {\bibinfo  {journal} {J.~Informetr.}\ }\textbf {\bibinfo {volume}
  {13}},\ \bibinfo {pages} {726} (\bibinfo {year} {2019})}\BibitemShut
  {NoStop}%
\bibitem [{\citenamefont {Golosovsky}\ and\ \citenamefont
  {Solomon}(2017)}]{Golosovsky17}%
  \BibitemOpen
  \bibfield  {author} {\bibinfo {author} {\bibfnamefont {M.}~\bibnamefont
  {Golosovsky}}\ and\ \bibinfo {author} {\bibfnamefont {S.}~\bibnamefont
  {Solomon}},\ }\href {https://doi.org/10.1103/PhysRevE.95.012324} {\bibfield
  {journal} {\bibinfo  {journal} {Phys.~Rev.~E}\ }\textbf {\bibinfo {volume}
  {95}},\ \bibinfo {pages} {012324} (\bibinfo {year} {2017})}\BibitemShut
  {NoStop}%
\bibitem [{\citenamefont {Okamura}(2022)}]{Okamura22}%
  \BibitemOpen
  \bibfield  {author} {\bibinfo {author} {\bibfnamefont {K.}~\bibnamefont
  {Okamura}},\ }\href {https://doi.org/10.1162/qss_a_00174} {\bibfield
  {journal} {\bibinfo  {journal} {Quant.~Sci.~Stud.}\ }\textbf {\bibinfo
  {volume} {3}},\ \bibinfo {pages} {122} (\bibinfo {year} {2022})}\BibitemShut
  {NoStop}%
\bibitem [{\citenamefont {Thelwall}(2016)}]{Thelwall16b}%
  \BibitemOpen
  \bibfield  {author} {\bibinfo {author} {\bibfnamefont {M.}~\bibnamefont
  {Thelwall}},\ }\href {https://doi.org/10.1016/j.joi.2016.07.006} {\bibfield
  {journal} {\bibinfo  {journal} {J.~Informetr.}\ }\textbf {\bibinfo {volume}
  {10}},\ \bibinfo {pages} {863} (\bibinfo {year} {2016})}\BibitemShut
  {NoStop}%
\bibitem [{\citenamefont {Waltman}\ \emph {et~al.}(2012)\citenamefont
  {Waltman}, \citenamefont {van Eck},\ and\ \citenamefont {van
  Raan}}]{Waltman12}%
  \BibitemOpen
  \bibfield  {author} {\bibinfo {author} {\bibfnamefont {L.}~\bibnamefont
  {Waltman}}, \bibinfo {author} {\bibfnamefont {N.~J.}\ \bibnamefont {van
  Eck}},\ and\ \bibinfo {author} {\bibfnamefont {A.~F.~J.}\ \bibnamefont {van
  Raan}},\ }\href {https://doi.org/10.1002/asi.21671} {\bibfield  {journal}
  {\bibinfo  {journal} {J.~Am.~Soc.~Inf.~Sci.~Tech.}\ }\textbf {\bibinfo
  {volume} {63}},\ \bibinfo {pages} {72} (\bibinfo {year} {2012})}\BibitemShut
  {NoStop}%
\bibitem [{\citenamefont {Golosovsky}(2021)}]{Golosovsky21}%
  \BibitemOpen
  \bibfield  {author} {\bibinfo {author} {\bibfnamefont {M.}~\bibnamefont
  {Golosovsky}},\ }\href {https://doi.org/10.1162/qss_a_00127} {\bibfield
  {journal} {\bibinfo  {journal} {Quant.~Sci.~Stud.}\ }\textbf {\bibinfo
  {volume} {2}},\ \bibinfo {pages} {527} (\bibinfo {year} {2021})}\BibitemShut
  {NoStop}%
\bibitem [{\citenamefont {Line}\ and\ \citenamefont {Sandison}(1974)}]{Line74}%
  \BibitemOpen
  \bibfield  {author} {\bibinfo {author} {\bibfnamefont {M.~B.}\ \bibnamefont
  {Line}}\ and\ \bibinfo {author} {\bibfnamefont {A.}~\bibnamefont
  {Sandison}},\ }\href {https://doi.org/10.1108/eb026583} {\bibfield  {journal}
  {\bibinfo  {journal} {J.~Doc.}\ }\textbf {\bibinfo {volume} {30}},\ \bibinfo
  {pages} {283} (\bibinfo {year} {1974})}\BibitemShut {NoStop}%
\bibitem [{\citenamefont {Nakamoto}(1988)}]{Nakamoto88}%
  \BibitemOpen
  \bibfield  {author} {\bibinfo {author} {\bibfnamefont {H.}~\bibnamefont
  {Nakamoto}},\ }\href
  {https://documentserver.uhasselt.be/bitstream/1942/837/1/nakamoto157.pdf}
  {\bibfield  {journal} {\bibinfo  {journal} {Informetrics}\ }\textbf {\bibinfo
  {volume} {87/88}},\ \bibinfo {pages} {157} (\bibinfo {year}
  {1988})}\BibitemShut {NoStop}%
\bibitem [{\citenamefont {Larivi\`{e}re}\ \emph {et~al.}(2008)\citenamefont
  {Larivi\`{e}re}, \citenamefont {Archambault},\ and\ \citenamefont
  {Gingras}}]{Lariviere08}%
  \BibitemOpen
  \bibfield  {author} {\bibinfo {author} {\bibfnamefont {V.}~\bibnamefont
  {Larivi\`{e}re}}, \bibinfo {author} {\bibfnamefont {E.}~\bibnamefont
  {Archambault}},\ and\ \bibinfo {author} {\bibfnamefont {Y.}~\bibnamefont
  {Gingras}},\ }\href {https://doi.org/10.1002/asi.20744} {\bibfield  {journal}
  {\bibinfo  {journal} {J.~Am.~Soc.~Inf.~Sci.~Tech.}\ }\textbf {\bibinfo
  {volume} {59}},\ \bibinfo {pages} {288} (\bibinfo {year} {2008})}\BibitemShut
  {NoStop}%
\bibitem [{\citenamefont {Parolo}\ \emph {et~al.}(2015)\citenamefont {Parolo},
  \citenamefont {Pan}, \citenamefont {Ghosh}, \citenamefont {Huberman},
  \citenamefont {Kaski},\ and\ \citenamefont {Fortunato}}]{Parolo15}%
  \BibitemOpen
  \bibfield  {author} {\bibinfo {author} {\bibfnamefont {P.~D.~B.}\
  \bibnamefont {Parolo}}, \bibinfo {author} {\bibfnamefont {R.~K.}\
  \bibnamefont {Pan}}, \bibinfo {author} {\bibfnamefont {R.}~\bibnamefont
  {Ghosh}}, \bibinfo {author} {\bibfnamefont {B.~A.}\ \bibnamefont {Huberman}},
  \bibinfo {author} {\bibfnamefont {K.}~\bibnamefont {Kaski}},\ and\ \bibinfo
  {author} {\bibfnamefont {S.}~\bibnamefont {Fortunato}},\ }\href
  {https://doi.org/10.1016/j.joi.2015.07.006} {\bibfield  {journal} {\bibinfo
  {journal} {J.~Informetr.}\ }\textbf {\bibinfo {volume} {9}},\ \bibinfo
  {pages} {734} (\bibinfo {year} {2015})}\BibitemShut {NoStop}%
\bibitem [{\citenamefont {Yin}\ and\ \citenamefont {Wang}(2017)}]{Yin17}%
  \BibitemOpen
  \bibfield  {author} {\bibinfo {author} {\bibfnamefont {Y.}~\bibnamefont
  {Yin}}\ and\ \bibinfo {author} {\bibfnamefont {D.}~\bibnamefont {Wang}},\
  }\href {https://doi.org/10.1016/j.joi.2017.04.002} {\bibfield  {journal}
  {\bibinfo  {journal} {J.~Informetr.}\ }\textbf {\bibinfo {volume} {11}},\
  \bibinfo {pages} {608} (\bibinfo {year} {2017})}\BibitemShut {NoStop}%
\bibitem [{\citenamefont {Golosovsky}\ and\ \citenamefont
  {Solomon}(2012{\natexlab{a}})}]{Golosovsky12b}%
  \BibitemOpen
  \bibfield  {author} {\bibinfo {author} {\bibfnamefont {M.}~\bibnamefont
  {Golosovsky}}\ and\ \bibinfo {author} {\bibfnamefont {S.}~\bibnamefont
  {Solomon}},\ }\href {https://doi.org/10.1140/epjst/e2012-01576-4} {\bibfield
  {journal} {\bibinfo  {journal} {Eur.~Phys.~J.~Spec.~Top.}\ }\textbf {\bibinfo
  {volume} {205}},\ \bibinfo {pages} {303} (\bibinfo {year}
  {2012}{\natexlab{a}})}\BibitemShut {NoStop}%
\bibitem [{\citenamefont {Golosovsky}\ and\ \citenamefont
  {Solomon}(2012{\natexlab{b}})}]{Golosovsky12}%
  \BibitemOpen
  \bibfield  {author} {\bibinfo {author} {\bibfnamefont {M.}~\bibnamefont
  {Golosovsky}}\ and\ \bibinfo {author} {\bibfnamefont {S.}~\bibnamefont
  {Solomon}},\ }\href {https://doi.org/10.1103/PhysRevLett.109.098701}
  {\bibfield  {journal} {\bibinfo  {journal} {Phys.~Rev.~Lett.}\ }\textbf
  {\bibinfo {volume} {109}},\ \bibinfo {pages} {098701} (\bibinfo {year}
  {2012}{\natexlab{b}})}\BibitemShut {NoStop}%
\bibitem [{\citenamefont {Wang}\ \emph {et~al.}(2013)\citenamefont {Wang},
  \citenamefont {Song},\ and\ \citenamefont {Barab\'asi}}]{Wang13}%
  \BibitemOpen
  \bibfield  {author} {\bibinfo {author} {\bibfnamefont {D.}~\bibnamefont
  {Wang}}, \bibinfo {author} {\bibfnamefont {C.}~\bibnamefont {Song}},\ and\
  \bibinfo {author} {\bibfnamefont {A.-L.}\ \bibnamefont {Barab\'asi}},\ }\href
  {https://doi.org/10.1126/science.1237825} {\bibfield  {journal} {\bibinfo
  {journal} {Science}\ }\textbf {\bibinfo {volume} {342}},\ \bibinfo {pages}
  {127} (\bibinfo {year} {2013})}\BibitemShut {NoStop}%
\bibitem [{\citenamefont {Seglen}(1992)}]{Seglen92}%
  \BibitemOpen
  \bibfield  {author} {\bibinfo {author} {\bibfnamefont {P.~O.}\ \bibnamefont
  {Seglen}},\ }\href
  {https://doi.org//10.1002/(SICI)1097-4571(199210)43:9<628::AID-ASI5>3.0.CO;2-0}
  {\bibfield  {journal} {\bibinfo  {journal} {J.~Am.~Soc.~Inf.~Sci.}\ }\textbf
  {\bibinfo {volume} {43}},\ \bibinfo {pages} {628} (\bibinfo {year}
  {1992})}\BibitemShut {NoStop}%
\bibitem [{\citenamefont {van Raan}(2004)}]{Raan04}%
  \BibitemOpen
  \bibfield  {author} {\bibinfo {author} {\bibfnamefont {A.~F.~J.}\
  \bibnamefont {van Raan}},\ }\href
  {https://doi.org/10.1023/B:SCIE.0000018543.82441.f1} {\bibfield  {journal}
  {\bibinfo  {journal} {Scientometrics}\ }\textbf {\bibinfo {volume} {59}},\
  \bibinfo {pages} {467} (\bibinfo {year} {2004})}\BibitemShut {NoStop}%
\bibitem [{\citenamefont {Ke}\ \emph {et~al.}(2015)\citenamefont {Ke},
  \citenamefont {Ferrara}, \citenamefont {Radicchi},\ and\ \citenamefont
  {Flammini}}]{Ke15}%
  \BibitemOpen
  \bibfield  {author} {\bibinfo {author} {\bibfnamefont {Q.}~\bibnamefont
  {Ke}}, \bibinfo {author} {\bibfnamefont {E.}~\bibnamefont {Ferrara}},
  \bibinfo {author} {\bibfnamefont {F.}~\bibnamefont {Radicchi}},\ and\
  \bibinfo {author} {\bibfnamefont {A.}~\bibnamefont {Flammini}},\ }\href
  {https://doi.org/10.1073/pnas.1424329112} {\bibfield  {journal} {\bibinfo
  {journal} {Proc.~Natl.~Acad.~Sci.~USA}\ }\textbf {\bibinfo {volume} {112}},\
  \bibinfo {pages} {7426} (\bibinfo {year} {2015})}\BibitemShut {NoStop}%
\bibitem [{\citenamefont {Barab\'{a}si}\ and\ \citenamefont
  {Albert}(1999)}]{Barabasi99}%
  \BibitemOpen
  \bibfield  {author} {\bibinfo {author} {\bibfnamefont {A.-L.}\ \bibnamefont
  {Barab\'{a}si}}\ and\ \bibinfo {author} {\bibfnamefont {R.}~\bibnamefont
  {Albert}},\ }\href {https://doi.org/10.1126/science.286.5439.50} {\bibfield
  {journal} {\bibinfo  {journal} {Science}\ }\textbf {\bibinfo {volume}
  {286}},\ \bibinfo {pages} {509} (\bibinfo {year} {1999})}\BibitemShut
  {NoStop}%
\bibitem [{\citenamefont {Barab\'{a}si}\ \emph {et~al.}(2002)\citenamefont
  {Barab\'{a}si}, \citenamefont {Jeong}, \citenamefont {N\'{e}da}, \citenamefont
  {Ravasz}, \citenamefont {Schubert},\ and\ \citenamefont
  {Vicsek}}]{Barabasi02}%
  \BibitemOpen
  \bibfield  {author} {\bibinfo {author} {\bibfnamefont {A.-L.}\ \bibnamefont
  {Barab\'{a}si}}, \bibinfo {author} {\bibfnamefont {H.}~\bibnamefont {Jeong}},
  \bibinfo {author} {\bibfnamefont {Z.}~\bibnamefont {N\'{e}da}}, \bibinfo
  {author} {\bibfnamefont {E.}~\bibnamefont {Ravasz}}, \bibinfo {author}
  {\bibfnamefont {A.}~\bibnamefont {Schubert}},\ and\ \bibinfo {author}
  {\bibfnamefont {T.}~\bibnamefont {Vicsek}},\ }\href
  {https://doi.org/10.1016/S0378-4371(02)00736-7} {\bibfield  {journal}
  {\bibinfo  {journal} {Physica A}\ }\textbf {\bibinfo {volume} {311}},\
  \bibinfo {pages} {590} (\bibinfo {year} {2002})}\BibitemShut {NoStop}%
\bibitem [{\citenamefont {Jeong}\ \emph {et~al.}(2003)\citenamefont {Jeong},
  \citenamefont {N\'{e}da},\ and\ \citenamefont {Barab\'{a}si}}]{Jeong03}%
  \BibitemOpen
  \bibfield  {author} {\bibinfo {author} {\bibfnamefont {H.}~\bibnamefont
  {Jeong}}, \bibinfo {author} {\bibfnamefont {Z.}~\bibnamefont {N\'{e}da}},\
  and\ \bibinfo {author} {\bibfnamefont {A.-L.}\ \bibnamefont {Barab\'{a}si}},\
  }\href {https://doi.org/10.1209/epl/i2003-00166-9} {\bibfield  {journal}
  {\bibinfo  {journal} {Europhys.~Lett.}\ }\textbf {\bibinfo {volume} {61}},\
  \bibinfo {pages} {567} (\bibinfo {year} {2003})}\BibitemShut {NoStop}%
\bibitem [{\citenamefont {Newman}(2009)}]{Newman09}%
  \BibitemOpen
  \bibfield  {author} {\bibinfo {author} {\bibfnamefont {M.~E.~J.}\
  \bibnamefont {Newman}},\ }\href {https://doi.org/10.1209/0295-5075/86/68001}
  {\bibfield  {journal} {\bibinfo  {journal} {Europhys.~Lett.}\ }\textbf
  {\bibinfo {volume} {86}},\ \bibinfo {pages} {68001} (\bibinfo {year}
  {2009})}\BibitemShut {NoStop}%
\bibitem [{\citenamefont {Sutton}(1997)}]{Sutton97}%
  \BibitemOpen
  \bibfield  {author} {\bibinfo {author} {\bibfnamefont {J.}~\bibnamefont
  {Sutton}},\ }\href {https://www.jstor.org/stable/2729692} {\bibfield
  {journal} {\bibinfo  {journal} {J.~Econ.~Lit.}\ }\textbf {\bibinfo {volume}
  {35}},\ \bibinfo {pages} {40} (\bibinfo {year} {1997})}\BibitemShut {NoStop}%
\bibitem [{\citenamefont {Yule}(1925)}]{Yule25}%
  \BibitemOpen
  \bibfield  {author} {\bibinfo {author} {\bibfnamefont {G.~U.}\ \bibnamefont
  {Yule}},\ }\href {https://doi.org/10.1098/rstb.1925.0002} {\bibfield
  {journal} {\bibinfo  {journal} {Philos.~Trans.~R.~Soc.~Lond.~B}\ }\textbf
  {\bibinfo {volume} {213}},\ \bibinfo {pages} {21} (\bibinfo {year}
  {1925})}\BibitemShut {NoStop}%
\bibitem [{\citenamefont {Simon}(1955)}]{Simon55}%
  \BibitemOpen
  \bibfield  {author} {\bibinfo {author} {\bibfnamefont {H.~A.}\ \bibnamefont
  {Simon}},\ }\href {https://doi.org/10.1093/biomet/42.3-4.425} {\bibfield
  {journal} {\bibinfo  {journal} {Biometrika}\ }\textbf {\bibinfo {volume}
  {42}},\ \bibinfo {pages} {425} (\bibinfo {year} {1955})}\BibitemShut
  {NoStop}%
\bibitem [{\citenamefont {Uzzi}\ \emph {et~al.}(2013)\citenamefont {Uzzi},
  \citenamefont {Mukherjee}, \citenamefont {Stringer},\ and\ \citenamefont
  {Jones}}]{Uzzi13}%
  \BibitemOpen
  \bibfield  {author} {\bibinfo {author} {\bibfnamefont {B.}~\bibnamefont
  {Uzzi}}, \bibinfo {author} {\bibfnamefont {S.}~\bibnamefont {Mukherjee}},
  \bibinfo {author} {\bibfnamefont {M.}~\bibnamefont {Stringer}},\ and\
  \bibinfo {author} {\bibfnamefont {B.}~\bibnamefont {Jones}},\ }\href
  {https://doi.org/10.1126/science.1240474} {\bibfield  {journal} {\bibinfo
  {journal} {Science}\ }\textbf {\bibinfo {volume} {342}},\ \bibinfo {pages}
  {468} (\bibinfo {year} {2013})}\BibitemShut {NoStop}%
\bibitem [{\citenamefont {Stegehuis}\ \emph {et~al.}(2015)\citenamefont
  {Stegehuis}, \citenamefont {Litvak},\ and\ \citenamefont
  {Waltman}}]{Stegehuis15}%
  \BibitemOpen
  \bibfield  {author} {\bibinfo {author} {\bibfnamefont {C.}~\bibnamefont
  {Stegehuis}}, \bibinfo {author} {\bibfnamefont {N.}~\bibnamefont {Litvak}},\
  and\ \bibinfo {author} {\bibfnamefont {L.}~\bibnamefont {Waltman}},\ }\href
  {https://doi.org/10.1016/j.joi.2015.06.005} {\bibfield  {journal} {\bibinfo
  {journal} {J.~Informetr.}\ }\textbf {\bibinfo {volume} {9}},\ \bibinfo
  {pages} {642} (\bibinfo {year} {2015})}\BibitemShut {NoStop}%
\bibitem [{\citenamefont {Cao}\ \emph {et~al.}(2016)\citenamefont {Cao},
  \citenamefont {Chen},\ and\ \citenamefont {{Ray Liu}}}]{Cao16}%
  \BibitemOpen
  \bibfield  {author} {\bibinfo {author} {\bibfnamefont {X.}~\bibnamefont
  {Cao}}, \bibinfo {author} {\bibfnamefont {Y.}~\bibnamefont {Chen}},\ and\
  \bibinfo {author} {\bibfnamefont {K.}~\bibnamefont {{Ray Liu}}},\ }\href
  {https://doi.org/10.1016/j.joi.2016.02.006} {\bibfield  {journal} {\bibinfo
  {journal} {J.~Informetr.}\ }\textbf {\bibinfo {volume} {10}},\ \bibinfo
  {pages} {471} (\bibinfo {year} {2016})}\BibitemShut {NoStop}%
\bibitem [{\citenamefont {Gl\"{a}nzel}\ and\ \citenamefont
  {Schubert}(1995)}]{Glanzel95b}%
  \BibitemOpen
  \bibfield  {author} {\bibinfo {author} {\bibfnamefont {W.}~\bibnamefont
  {Gl\"{a}nzel}}\ and\ \bibinfo {author} {\bibfnamefont {A.}~\bibnamefont
  {Schubert}},\ }\href {https://doi.org/10.1016/0306-4573(95)80007-G}
  {\bibfield  {journal} {\bibinfo  {journal} {Inf.~Process.~Manag.}\ }\textbf
  {\bibinfo {volume} {31}},\ \bibinfo {pages} {69} (\bibinfo {year}
  {1995})}\BibitemShut {NoStop}%
\bibitem [{\citenamefont {Klemm}\ and\ \citenamefont
  {Egu\'{\i}luz}(2002)}]{Klemm02}%
  \BibitemOpen
  \bibfield  {author} {\bibinfo {author} {\bibfnamefont {K.}~\bibnamefont
  {Klemm}}\ and\ \bibinfo {author} {\bibfnamefont {V.~M.}\ \bibnamefont
  {Egu\'{\i}luz}},\ }\href {https://doi.org/10.1103/PhysRevE.65.036123}
  {\bibfield  {journal} {\bibinfo  {journal} {Phys.~Rev.~E}\ }\textbf {\bibinfo
  {volume} {65}},\ \bibinfo {pages} {036123} (\bibinfo {year}
  {2002})}\BibitemShut {NoStop}%
\bibitem [{\citenamefont {Tasaki}(2005)}]{Tasaki05}%
  \BibitemOpen
  \bibfield  {author} {\bibinfo {author} {\bibfnamefont {H.}~\bibnamefont
  {Tasaki}},\ }\bibinfo {title} {Brownian motion and nonequilibrium statistical
  mechanics},\ in\ \href {https://www.gakushuin.ac.jp/~881791/docs/BMNESM.pdf}
  {\emph {\bibinfo {booktitle} {Einstein and 21st-Century Physics}}},\ \bibinfo
  {editor} {edited by\ \bibinfo {editor} {\bibnamefont {{The Physical Society
  of Japan}}}}\ (\bibinfo  {publisher} {Nippon Hyoron Sha Co., Ltd.},\ \bibinfo
  {address} {Tokyo},\ \bibinfo
  {year} {2005})\BibitemShut {NoStop}%
\bibitem [{\citenamefont {Black}\ and\ \citenamefont
  {Scholes}(1973)}]{Black73}%
  \BibitemOpen
  \bibfield  {author} {\bibinfo {author} {\bibfnamefont {F.}~\bibnamefont
  {Black}}\ and\ \bibinfo {author} {\bibfnamefont {M.}~\bibnamefont
  {Scholes}},\ }\href {https://www.jstor.org/stable/1831029} {\bibfield
  {journal} {\bibinfo  {journal} {J.~Political Econ.}\ }\textbf {\bibinfo
  {volume} {81}},\ \bibinfo {pages} {637} (\bibinfo {year} {1973})}\BibitemShut
  {NoStop}%
\bibitem [{\citenamefont {Merton}(1973)}]{Merton73}%
  \BibitemOpen
  \bibfield  {author} {\bibinfo {author} {\bibfnamefont {R.~C.}\ \bibnamefont
  {Merton}},\ }\href {https://doi.org/10.2307/3003143} {\bibfield  {journal}
  {\bibinfo  {journal} {Bell J.~Econ.~Manag.~Sci.}\ }\textbf {\bibinfo {volume}
  {4}},\ \bibinfo {pages} {141} (\bibinfo {year} {1973})}\BibitemShut {NoStop}%
\bibitem [{sup()}]{suppletext}%
  \BibitemOpen
  \href@noop {} {}\bibinfo {note} {See supplemental material at [URL will be inserted by publisher] for supplementary text (Secs.~II--VI), Figs.~S1--S6 and Tables~S1--S2.}\BibitemShut
  {Stop}%
\bibitem [{\citenamefont {arXiv}(2020)}]{arXiv}%
  \BibitemOpen
  \bibfield  {author} \href {https://arxiv.org/}{\bibinfo {title} {arXiv.org}},\ \bibinfo {note}
  {accessed January, 2020}\BibitemShut {NoStop}%
\bibitem [{\citenamefont {Okamura}\ and\ \citenamefont
  {Koshiba}(2021)}]{Okamura21}%
  \BibitemOpen
  \bibfield  {author} {\bibinfo {author} {\bibfnamefont {K.}~\bibnamefont
  {Okamura}}\ and\ \bibinfo {author} {\bibfnamefont {H.}~\bibnamefont
  {Koshiba}},\ }\href {https://doi.org/10.5281/zenodo.5803962} {\bibinfo
  {title} {Citation data of ar{X}iv eprints and the associated
  quantitatively-and-temporally normalised impact metrics}},\ \bibinfo
  {howpublished} {[Data set]} (\bibinfo {year} {2021}),\ \bibinfo {note}
  {{Zenodo}}\BibitemShut {NoStop}%
\bibitem [{\citenamefont {Mandelbrot}\ and\ \citenamefont
  {Van~Ness}(1968)}]{Mandelbrot68}%
  \BibitemOpen
  \bibfield  {author} {\bibinfo {author} {\bibfnamefont {B.~B.}\ \bibnamefont
  {Mandelbrot}}\ and\ \bibinfo {author} {\bibfnamefont {J.~W.}\ \bibnamefont
  {Van~Ness}},\ }\href {https://doi.org/10.1137/1010093} {\bibfield  {journal}
  {\bibinfo  {journal} {SIAM Rev.}\ }\textbf {\bibinfo {volume} {10}},\
  \bibinfo {pages} {422} (\bibinfo {year} {1968})}\BibitemShut {NoStop}%
\bibitem [{\citenamefont {Pipiras}\ and\ \citenamefont
  {Taqqu}(2017)}]{Pipiras17}%
  \BibitemOpen
  \bibfield  {author} {\bibinfo {author} {\bibfnamefont {V.}~\bibnamefont
  {Pipiras}}\ and\ \bibinfo {author} {\bibfnamefont {M.~S.}\ \bibnamefont
  {Taqqu}},\ }\href {https://doi.org/10.1017/CBO9781139600347} {\emph {\bibinfo
  {title} {Long-Range Dependence and Self-Similarity}}},\ Cambridge Series in
  Statistical and Probabilistic Mathematics\ (\bibinfo  {publisher} {Cambridge
  University Press},\ \bibinfo
  {address} {Cambridge, UK},\ \bibinfo {year} {2017})\BibitemShut {NoStop}%
\bibitem [{\citenamefont {Allegrini}\ \emph {et~al.}(1998)\citenamefont
  {Allegrini}, \citenamefont {Buiatti}, \citenamefont {Grigolini},\ and\
  \citenamefont {West}}]{Allegrini98}%
  \BibitemOpen
  \bibfield  {author} {\bibinfo {author} {\bibfnamefont {P.}~\bibnamefont
  {Allegrini}}, \bibinfo {author} {\bibfnamefont {M.}~\bibnamefont {Buiatti}},
  \bibinfo {author} {\bibfnamefont {P.}~\bibnamefont {Grigolini}},\ and\
  \bibinfo {author} {\bibfnamefont {B.~J.}\ \bibnamefont {West}},\ }\href
  {https://doi.org/10.1103/PhysRevE.57.4558} {\bibfield  {journal} {\bibinfo
  {journal} {Phys.~Rev.~E}\ }\textbf {\bibinfo {volume} {57}},\ \bibinfo
  {pages} {4558} (\bibinfo {year} {1998})}\BibitemShut {NoStop}%
\bibitem [{\citenamefont {Jeon}\ \emph {et~al.}(2011)\citenamefont {Jeon},
  \citenamefont {Tejedor}, \citenamefont {Burov}, \citenamefont {Barkai},
  \citenamefont {Selhuber-Unkel}, \citenamefont {Berg-S\o{}rensen},
  \citenamefont {Oddershede},\ and\ \citenamefont {Metzler}}]{Jeon11}%
  \BibitemOpen
  \bibfield  {author} {\bibinfo {author} {\bibfnamefont {J.-H.}\ \bibnamefont
  {Jeon}}, \bibinfo {author} {\bibfnamefont {V.}~\bibnamefont {Tejedor}},
  \bibinfo {author} {\bibfnamefont {S.}~\bibnamefont {Burov}}, \bibinfo
  {author} {\bibfnamefont {E.}~\bibnamefont {Barkai}}, \bibinfo {author}
  {\bibfnamefont {C.}~\bibnamefont {Selhuber-Unkel}}, \bibinfo {author}
  {\bibfnamefont {K.}~\bibnamefont {Berg-S\o{}rensen}}, \bibinfo {author}
  {\bibfnamefont {L.}~\bibnamefont {Oddershede}},\ and\ \bibinfo {author}
  {\bibfnamefont {R.}~\bibnamefont {Metzler}},\ }\href
  {https://doi.org/10.1103/PhysRevLett.106.048103} {\bibfield  {journal}
  {\bibinfo  {journal} {Phys.~Rev.~Lett.}\ }\textbf {\bibinfo {volume} {106}},\
  \bibinfo {pages} {048103} (\bibinfo {year} {2011})}\BibitemShut {NoStop}%
\bibitem [{\citenamefont {Molz}\ \emph {et~al.}(1997)\citenamefont {Molz},
  \citenamefont {Liu},\ and\ \citenamefont {Szulga}}]{Molz97}%
  \BibitemOpen
  \bibfield  {author} {\bibinfo {author} {\bibfnamefont {F.~J.}\ \bibnamefont
  {Molz}}, \bibinfo {author} {\bibfnamefont {H.~H.}\ \bibnamefont {Liu}},\ and\
  \bibinfo {author} {\bibfnamefont {J.}~\bibnamefont {Szulga}},\ }\href
  {https://doi.org/10.1029/97WR01982} {\bibfield  {journal} {\bibinfo
  {journal} {Water Resour.~Res.}\ }\textbf {\bibinfo {volume} {33}},\ \bibinfo
  {pages} {2273} (\bibinfo {year} {1997})}\BibitemShut {NoStop}%
\bibitem [{\citenamefont {Leland}\ \emph {et~al.}(1994)\citenamefont {Leland},
  \citenamefont {Taqqu}, \citenamefont {Willinger},\ and\ \citenamefont
  {Wilson}}]{Leland94}%
  \BibitemOpen
  \bibfield  {author} {\bibinfo {author} {\bibfnamefont {W.~E.}\ \bibnamefont
  {Leland}}, \bibinfo {author} {\bibfnamefont {M.~S.}\ \bibnamefont {Taqqu}},
  \bibinfo {author} {\bibfnamefont {W.}~\bibnamefont {Willinger}},\ and\
  \bibinfo {author} {\bibfnamefont {D.~V.}\ \bibnamefont {Wilson}},\ }\href
  {https://doi.org/10.1109/90.282603} {\bibfield  {journal} {\bibinfo
  {journal} {IEEE/ACM Trans.~Netw.}\ }\textbf {\bibinfo {volume} {2}},\
  \bibinfo {pages} {1} (\bibinfo {year} {1994})}\BibitemShut {NoStop}%
\bibitem [{\citenamefont {Abry}\ \emph {et~al.}(2000)\citenamefont {Abry},
  \citenamefont {Flandrin}, \citenamefont {Taqqu},\ and\ \citenamefont
  {Veitch}}]{Abry00}%
  \BibitemOpen
  \bibfield  {author} {\bibinfo {author} {\bibfnamefont {P.}~\bibnamefont
  {Abry}}, \bibinfo {author} {\bibfnamefont {P.}~\bibnamefont {Flandrin}},
  \bibinfo {author} {\bibfnamefont {M.~S.}\ \bibnamefont {Taqqu}},\ and\
  \bibinfo {author} {\bibfnamefont {D.}~\bibnamefont {Veitch}},\ }\bibinfo
  {title} {Wavelets for the analysis, estimation, and synthesis of scaling
  data},\ in\ \href {https://doi.org/10.1002/047120644X.ch2} {\emph {\bibinfo
  {booktitle} {Self-Similar Network Traffic and Performance Evaluation}}}\
  (\bibinfo  {publisher} {John Wiley \& Sons, Ltd},\ \bibinfo
  {address} {New York},\ \bibinfo {year} {2000}),\
  pp.\ \bibinfo {pages} {39--88}\BibitemShut
  {NoStop}%
\bibitem [{\citenamefont {Mikosch}\ \emph {et~al.}(2002)\citenamefont
  {Mikosch}, \citenamefont {Resnick}, \citenamefont {Rootz\'{e}n},\ and\
  \citenamefont {Stegeman}}]{Mikosch02}%
  \BibitemOpen
  \bibfield  {author} {\bibinfo {author} {\bibfnamefont {T.}~\bibnamefont
  {Mikosch}}, \bibinfo {author} {\bibfnamefont {S.}~\bibnamefont {Resnick}},
  \bibinfo {author} {\bibfnamefont {H.}~\bibnamefont {Rootz\'{e}n}},\ and\
  \bibinfo {author} {\bibfnamefont {A.}~\bibnamefont {Stegeman}},\ }\href
  {https://doi.org/10.1214/aoap/1015961155} {\bibfield  {journal} {\bibinfo
  {journal} {Ann.~Appl.~Probab.}\ }\textbf {\bibinfo {volume} {12}},\ \bibinfo
  {pages} {23} (\bibinfo {year} {2002})}\BibitemShut {NoStop}%
\bibitem [{\citenamefont {Cutland}\ \emph {et~al.}(1995)\citenamefont
  {Cutland}, \citenamefont {Kopp},\ and\ \citenamefont
  {Willinger}}]{Cutland95}%
  \BibitemOpen
  \bibfield  {author} {\bibinfo {author} {\bibfnamefont {N.~J.}\ \bibnamefont
  {Cutland}}, \bibinfo {author} {\bibfnamefont {P.~E.}\ \bibnamefont {Kopp}},\
  and\ \bibinfo {author} {\bibfnamefont {W.}~\bibnamefont {Willinger}},\ }in\
  \href {https://doi.org/10.1007/978-3-0348-7026-9_23} {\emph {\bibinfo
  {booktitle} {Seminar on Stochastic Analysis, Random Fields and
  Applications}}},\ \bibinfo {editor} {edited by\ \bibinfo {editor}
  {\bibfnamefont {E.}~\bibnamefont {Bolthausen}}, \bibinfo {editor}
  {\bibfnamefont {M.}~\bibnamefont {Dozzi}},\ and\ \bibinfo {editor}
  {\bibfnamefont {F.}~\bibnamefont {Russo}}}\ (\bibinfo  {publisher}
  {Birkh\"{a}user Basel},\ \bibinfo {year} {1995})\ pp.\ \bibinfo {pages}
  {327--351}\BibitemShut {NoStop}%
\bibitem [{\citenamefont {Comte}\ and\ \citenamefont
  {Renault}(1998)}]{Comte98}%
  \BibitemOpen
  \bibfield  {author} {\bibinfo {author} {\bibfnamefont {F.}~\bibnamefont
  {Comte}}\ and\ \bibinfo {author} {\bibfnamefont {E.}~\bibnamefont
  {Renault}},\ }\href {https://doi.org/10.1111/1467-9965.00057} {\bibfield
  {journal} {\bibinfo  {journal} {Math.~Finance}\ }\textbf {\bibinfo {volume}
  {8}},\ \bibinfo {pages} {291} (\bibinfo {year} {1998})}\BibitemShut {NoStop}%
\bibitem [{\citenamefont {Gatheral}\ \emph {et~al.}(2018)\citenamefont
  {Gatheral}, \citenamefont {Jaisson},\ and\ \citenamefont
  {Rosenbaum}}]{Gatheral18}%
  \BibitemOpen
  \bibfield  {author} {\bibinfo {author} {\bibfnamefont {J.}~\bibnamefont
  {Gatheral}}, \bibinfo {author} {\bibfnamefont {T.}~\bibnamefont {Jaisson}},\
  and\ \bibinfo {author} {\bibfnamefont {M.}~\bibnamefont {Rosenbaum}},\ }\href
  {https://doi.org/10.1080/14697688.2017.1393551} {\bibfield  {journal}
  {\bibinfo  {journal} {Quantit.~Finance}\ }\textbf {\bibinfo {volume} {18}},\
  \bibinfo {pages} {933} (\bibinfo {year} {2018})}\BibitemShut {NoStop}%
\bibitem [{\citenamefont {Hu}\ and\ \citenamefont {{\O}ksendal}(2003)}]{Hu03}%
  \BibitemOpen
  \bibfield  {author} {\bibinfo {author} {\bibfnamefont {Y.}~\bibnamefont
  {Hu}}\ and\ \bibinfo {author} {\bibfnamefont {B.}~\bibnamefont
  {{\O}ksendal}},\ }\href {https://doi.org/10.1142/S0219025703001110}
  {\bibfield  {journal} {\bibinfo  {journal} {Infin.~Dimens.~Anal.~Quantum
  Probab.~Relat.~Top.}\ }\textbf {\bibinfo {volume} {06}},\ \bibinfo {pages}
  {1} (\bibinfo {year} {2003})}\BibitemShut {NoStop}%
\bibitem [{\citenamefont {Biagini}\ \emph {et~al.}(2004)\citenamefont
  {Biagini}, \citenamefont {{\O}ksendal}, \citenamefont {Sulem},\ and\
  \citenamefont {Wallner}}]{Biagini04}%
  \BibitemOpen
  \bibfield  {author} {\bibinfo {author} {\bibfnamefont {F.}~\bibnamefont
  {Biagini}}, \bibinfo {author} {\bibfnamefont {B.}~\bibnamefont
  {{\O}ksendal}}, \bibinfo {author} {\bibfnamefont {A.}~\bibnamefont {Sulem}},\
  and\ \bibinfo {author} {\bibfnamefont {N.}~\bibnamefont {Wallner}},\ }\href
  {https://doi.org/10.1098/rspa.2003.1246} {\bibfield  {journal} {\bibinfo
  {journal} {Proc.~R.~Soc.~A}\ }\textbf {\bibinfo {volume} {460}},\ \bibinfo
  {pages} {347} (\bibinfo {year} {2004})}\BibitemShut {NoStop}%
\bibitem [{\citenamefont {Fenton}(1960)}]{Fenton60}%
  \BibitemOpen
  \bibfield  {author} {\bibinfo {author} {\bibfnamefont {L.}~\bibnamefont
  {Fenton}},\ }\href {https://doi.org/10.1109/TCOM.1960.1097606} {\bibfield
  {journal} {\bibinfo  {journal} {IRE Trans.~Commun.}\ }\textbf {\bibinfo
  {volume} {8}},\ \bibinfo {pages} {57} (\bibinfo {year} {1960})}\BibitemShut
  {NoStop}%
\bibitem [{\citenamefont {Flynn}(2019)}]{Flynn19}%
  \BibitemOpen
  \bibfield  {author} {\bibinfo {author} {\bibfnamefont {C.}~\bibnamefont
  {Flynn}},\ }\href {https://github.com/crflynn/fbm} {\bibinfo {title} {fbm:
  {E}xact methods for simulating fractional {B}rownian motion and fractional
  {G}aussian noise in {P}ython, {V}er.~0.3.0}},\ \bibinfo {howpublished}
  {[Computer software]} (\bibinfo {year} {2019}),\ \bibinfo {note}
  {{GitHub}}\BibitemShut {NoStop}%
\bibitem [{\citenamefont {semanticscholar}(2020)}]{semanticscholar}%
  \BibitemOpen
  \bibfield  {author} \href {https://api.semanticscholar.org/}{\bibinfo {title} 
  {Semantic Scholar API}},\ \bibinfo {note}
  {accessed January, 2020}\BibitemShut {NoStop}%
\bibitem [{\citenamefont {Hurst}(1951)}]{Hurst51}%
  \BibitemOpen
  \bibfield  {author} {\bibinfo {author} {\bibfnamefont {H.~E.}\ \bibnamefont
  {Hurst}},\ }\href {https://doi.org/10.1061/TACEAT.0006518} {\bibfield
  {journal} {\bibinfo  {journal} {Trans.~Am.~Soc.~Civil Eng.}\ }\textbf
  {\bibinfo {volume} {116}},\ \bibinfo {pages} {770} (\bibinfo {year}
  {1951})}\BibitemShut {NoStop}%
\bibitem [{\citenamefont {Mandelbrot}(1972)}]{Mandelbrot72}%
  \BibitemOpen
  \bibfield  {author} {\bibinfo {author} {\bibfnamefont {B.}~\bibnamefont
  {Mandelbrot}},\ }\bibinfo {title} {Statistical methodology for nonperiodic
  cycles: {F}rom the covariance to {R/S} analysis},\ in\ \href
  {http://www.nber.org/chapters/c9433} {\emph {\bibinfo {booktitle} {Annals of
  Economic and Social Measurement, Volume 1, number 3}}},\ \bibinfo {editor}
  {edited by\ \bibinfo {editor} {\bibfnamefont {S.~V.}\ \bibnamefont {Berg}}}\
  (\bibinfo  {publisher} {National Bureau of Economic Research, Inc},\ \bibinfo
  {address} {Washington, DC},\ \bibinfo
  {year} {1972})\ pp.\ \bibinfo {pages} {259--290}\BibitemShut {NoStop}%
\bibitem [{\citenamefont {Peng}\ \emph {et~al.}(1994)\citenamefont {Peng},
  \citenamefont {Buldyrev}, \citenamefont {Havlin}, \citenamefont {Simons},
  \citenamefont {Stanley},\ and\ \citenamefont {Goldberger}}]{Peng94}%
  \BibitemOpen
  \bibfield  {author} {\bibinfo {author} {\bibfnamefont {C.-K.}\ \bibnamefont
  {Peng}}, \bibinfo {author} {\bibfnamefont {S.~V.}\ \bibnamefont {Buldyrev}},
  \bibinfo {author} {\bibfnamefont {S.}~\bibnamefont {Havlin}}, \bibinfo
  {author} {\bibfnamefont {M.}~\bibnamefont {Simons}}, \bibinfo {author}
  {\bibfnamefont {H.~E.}\ \bibnamefont {Stanley}},\ and\ \bibinfo {author}
  {\bibfnamefont {A.~L.}\ \bibnamefont {Goldberger}},\ } \href
  {https://doi.org/10.1103/PhysRevE.49.1685} {\bibfield  {journal} {\bibinfo
  {journal} {Phys.~Rev.~E}\ }\textbf {\bibinfo {volume} {49}},\ \bibinfo
  {pages} {1685} (\bibinfo {year} {1994})}\BibitemShut {NoStop}%
\bibitem [{\citenamefont {Hu}\ \emph {et~al.}(2001)\citenamefont {Hu},
  \citenamefont {Ivanov}, \citenamefont {Chen}, \citenamefont {Carpena},\ and\
  \citenamefont {Stanley}}]{Hu01}%
  \BibitemOpen
  \bibfield  {author} {\bibinfo {author} {\bibfnamefont {K.}~\bibnamefont
  {Hu}}, \bibinfo {author} {\bibfnamefont {P.~C.}\ \bibnamefont {Ivanov}},
  \bibinfo {author} {\bibfnamefont {Z.}~\bibnamefont {Chen}}, \bibinfo {author}
  {\bibfnamefont {P.}~\bibnamefont {Carpena}},\ and\ \bibinfo {author}
  {\bibfnamefont {H.~E.}~\bibnamefont {Stanley}},\ } 
  \href {https://doi.org/10.1103/PhysRevE.64.011114} {\bibfield  {journal}
  {\bibinfo  {journal} {Phys.~Rev.~E}\ }\textbf {\bibinfo {volume} {64}},\
  \bibinfo {pages} {011114} (\bibinfo {year} {2001})}\BibitemShut {NoStop}%
\bibitem [{\citenamefont {Chen}\ \emph {et~al.}(2002)\citenamefont {Chen},
  \citenamefont {Ivanov}, \citenamefont {Hu},\ and\ \citenamefont
  {Stanley}}]{Chen02}%
  \BibitemOpen
  \bibfield  {author} {\bibinfo {author} {\bibfnamefont {Z.}~\bibnamefont
  {Chen}}, \bibinfo {author} {\bibfnamefont {P.~C.}\ \bibnamefont {Ivanov}},
  \bibinfo {author} {\bibfnamefont {K.}~\bibnamefont {Hu}},\ and\ \bibinfo
  {author} {\bibfnamefont {H.~E.}\ \bibnamefont {Stanley}},\ }
   \href {https://doi.org/10.1103/PhysRevE.65.041107}
  {\bibfield  {journal} {\bibinfo  {journal} {Phys.~Rev.~E}\ }\textbf {\bibinfo
  {volume} {65}},\ \bibinfo {pages} {041107} (\bibinfo {year}
  {2002})}\BibitemShut {NoStop}%
\bibitem [{\citenamefont {Batty}\ and\ \citenamefont
  {Longley}(1994)}]{Batty94}%
  \BibitemOpen
  \bibfield  {author} {\bibinfo {author} {\bibfnamefont {M.}~\bibnamefont
  {Batty}}\ and\ \bibinfo {author} {\bibfnamefont {P.}~\bibnamefont
  {Longley}},\ }\href {https://discovery.ucl.ac.uk/id/eprint/1370661/} {\emph
  {\bibinfo {title} {Fractal Cities: {A} Geometry of Form and Function}}}\
  (\bibinfo  {publisher} {Academic Press},\ \bibinfo
  {address} {San Diego, CA}\ \bibinfo {year}
  {1994})\BibitemShut {NoStop}%
\bibitem [{\citenamefont {Song}\ \emph {et~al.}(2005)\citenamefont {Song},
  \citenamefont {Havlin},\ and\ \citenamefont {Makse}}]{Song05}%
  \BibitemOpen
  \bibfield  {author} {\bibinfo {author} {\bibfnamefont {C.}~\bibnamefont
  {Song}}, \bibinfo {author} {\bibfnamefont {S.}~\bibnamefont {Havlin}},\ and\
  \bibinfo {author} {\bibfnamefont {H.~A.}\ \bibnamefont {Makse}},\ }\href
  {https://doi.org/10.1038/nature03248} {\bibfield  {journal} {\bibinfo
  {journal} {Nature}\ }\textbf {\bibinfo {volume} {433}},\ \bibinfo {pages}
  {392} (\bibinfo {year} {2005})}\BibitemShut {NoStop}%
\bibitem [{\citenamefont {Song}\ \emph {et~al.}(2006)\citenamefont {Song},
  \citenamefont {Havlin},\ and\ \citenamefont {Makse}}]{Song06}%
  \BibitemOpen
  \bibfield  {author} {\bibinfo {author} {\bibfnamefont {C.}~\bibnamefont
  {Song}}, \bibinfo {author} {\bibfnamefont {S.}~\bibnamefont {Havlin}},\ and\
  \bibinfo {author} {\bibfnamefont {H.~A.}\ \bibnamefont {Makse}},\ }\href
  {https://doi.org/10.1038/nphys266} {\bibfield  {journal} {\bibinfo  {journal}
  {Nat.~Phys.}\ }\textbf {\bibinfo {volume} {2}},\ \bibinfo {pages} {275}
  (\bibinfo {year} {2006})}\BibitemShut {NoStop}%
\bibitem [{\citenamefont {Fortunato}\ \emph {et~al.}(2018)\citenamefont
  {Fortunato}, \citenamefont {Bergstrom}, \citenamefont {B\"{o}rner},
  \citenamefont {Evans}, \citenamefont {Helbing}, \citenamefont
  {Milojevi\'{c}}, \citenamefont {Petersen}, \citenamefont {Radicchi},
  \citenamefont {Sinatra}, \citenamefont {Uzzi}, \citenamefont {Vespignani},
  \citenamefont {Waltman}, \citenamefont {Wang},\ and\ \citenamefont
  {Barab\'{a}si}}]{Fortunato18}%
  \BibitemOpen
  \bibfield  {author} {\bibinfo {author} {\bibfnamefont {S.}~\bibnamefont
  {Fortunato}}, \bibinfo {author} {\bibfnamefont {C.~T.}\ \bibnamefont
  {Bergstrom}}, \bibinfo {author} {\bibfnamefont {K.}~\bibnamefont
  {B\"{o}rner}}, \bibinfo {author} {\bibfnamefont {J.~A.}\ \bibnamefont
  {Evans}}, \bibinfo {author} {\bibfnamefont {D.}~\bibnamefont {Helbing}},
  \bibinfo {author} {\bibfnamefont {S.}~\bibnamefont {Milojevi\'{c}}}, \bibinfo
  {author} {\bibfnamefont {A.~M.}\ \bibnamefont {Petersen}}, \bibinfo {author}
  {\bibfnamefont {F.}~\bibnamefont {Radicchi}}, \bibinfo {author}
  {\bibfnamefont {R.}~\bibnamefont {Sinatra}}, \bibinfo {author} {\bibfnamefont
  {B.}~\bibnamefont {Uzzi}}, \bibinfo {author} {\bibfnamefont {A.}~\bibnamefont
  {Vespignani}}, \bibinfo {author} {\bibfnamefont {L.}~\bibnamefont {Waltman}},
  \bibinfo {author} {\bibfnamefont {D.}~\bibnamefont {Wang}},\ and\ \bibinfo
  {author} {\bibfnamefont {A.-L.}\ \bibnamefont {Barab\'{a}si}},\ }\href
  {https://doi.org/10.1126/science.aao0185} {\bibfield  {journal} {\bibinfo
  {journal} {Science}\ }\textbf {\bibinfo {volume} {359}},\ \bibinfo {pages}
  {eaao0185} (\bibinfo {year} {2018})}\BibitemShut {NoStop}%
\bibitem [{\citenamefont {Biagini}\ \emph {et~al.}(2008)\citenamefont
  {Biagini}, \citenamefont {Hu}, \citenamefont {{\O}ksendal},\ and\
  \citenamefont {Zhang}}]{Biagini08}%
  \BibitemOpen
  \bibfield  {author} {\bibinfo {author} {\bibfnamefont {F.}~\bibnamefont
  {Biagini}}, \bibinfo {author} {\bibfnamefont {Y.}~\bibnamefont {Hu}},
  \bibinfo {author} {\bibfnamefont {B.}~\bibnamefont {{\O}ksendal}},\ and\
  \bibinfo {author} {\bibfnamefont {T.}~\bibnamefont {Zhang}},\ }\href
  {https://doi.org/https://doi.org/10.1007/978-1-84628-797-8} {\emph {\bibinfo
  {title} {Stochastic Calculus for Fractional Brownian Motion and
  Applications}}}\ (\bibinfo  {publisher} {Springer London},\ \bibinfo {year}
  {2008})\BibitemShut {NoStop}%
\bibitem [{\citenamefont {Candia}\ \emph {et~al.}(2019)\citenamefont {Candia},
  \citenamefont {Jara-Figueroa}, \citenamefont {Rodriguez-Sickert},
  \citenamefont {Barab\'{a}si},\ and\ \citenamefont {Hidalgo}}]{Candia19}%
  \BibitemOpen
  \bibfield  {author} {\bibinfo {author} {\bibfnamefont {C.}~\bibnamefont
  {Candia}}, \bibinfo {author} {\bibfnamefont {C.}~\bibnamefont
  {Jara-Figueroa}}, \bibinfo {author} {\bibfnamefont {C.}~\bibnamefont
  {Rodriguez-Sickert}}, \bibinfo {author} {\bibfnamefont {A.-L.}\ \bibnamefont
  {Barab\'{a}si}},\ and\ \bibinfo {author} {\bibfnamefont {C.~A.}\ \bibnamefont
  {Hidalgo}},\ }\href {https://doi.org/10.1038/s41562-018-0474-5} {\bibfield
  {journal} {\bibinfo  {journal} {Nat.~Hum.~Behav.}\ }\textbf {\bibinfo
  {volume} {3}},\ \bibinfo {pages} {82} (\bibinfo {year} {2019})}\BibitemShut
  {NoStop}%
\bibitem [{\citenamefont {Lorenz-Spreen}\ \emph {et~al.}(2019)\citenamefont
  {Lorenz-Spreen}, \citenamefont {M{\o}nsted}, \citenamefont {H\"{o}vel},\ and\
  \citenamefont {Lehmann}}]{Lorenz-Spreen19}%
  \BibitemOpen
  \bibfield  {author} {\bibinfo {author} {\bibfnamefont {P.}~\bibnamefont
  {Lorenz-Spreen}}, \bibinfo {author} {\bibfnamefont {B.~M.}\ \bibnamefont
  {M{\o}nsted}}, \bibinfo {author} {\bibfnamefont {P.}~\bibnamefont
  {H\"{o}vel}},\ and\ \bibinfo {author} {\bibfnamefont {S.}~\bibnamefont
  {Lehmann}},\ }\href {https://doi.org/10.1038/s41467-019-09311-w} {\bibfield
  {journal} {\bibinfo  {journal} {Nat.~Commun.}\ }\textbf {\bibinfo {volume}
  {10}},\ \bibinfo {pages} {1759} (\bibinfo {year} {2019})}\BibitemShut
  {NoStop}%
\bibitem [{\citenamefont {Cheridito}(2001)}]{Cheridito01}%
  \BibitemOpen
  \bibfield  {author} {\bibinfo {author} {\bibfnamefont {P.}~\bibnamefont
  {Cheridito}},\ }\href {https://doi.org/10.2307/3318626} {\bibfield  {journal}
  {\bibinfo  {journal} {Bernoulli}\ }\textbf {\bibinfo {volume} {7}},\ \bibinfo
  {pages} {913} (\bibinfo {year} {2001})}\BibitemShut {NoStop}%
\bibitem [{\citenamefont {Th\"{a}le}(2009)}]{Thale09}%
  \BibitemOpen
  \bibfield  {author} {\bibinfo {author} {\bibfnamefont {C.}~\bibnamefont
  {Th\"{a}le}},\ }\href
  {https://folia.unifr.ch/documents/301263/files/thale_fbm.pdf} {\bibfield
  {journal} {\bibinfo  {journal} {Appl.~Math.~Sci.}\ }\textbf {\bibinfo
  {volume} {3}},\ \bibinfo {pages} {1885} (\bibinfo {year} {2009})}\BibitemShut
  {NoStop}%
\bibitem [{\citenamefont {Stojkoski}\ \emph {et~al.}(2021)\citenamefont
  {Stojkoski}, \citenamefont {Sandev}, \citenamefont {Kocarev},\ and\
  \citenamefont {Pal}}]{Stojkoski21}%
  \BibitemOpen
  \bibfield  {author} {\bibinfo {author} {\bibfnamefont {V.}~\bibnamefont
  {Stojkoski}}, \bibinfo {author} {\bibfnamefont {T.}~\bibnamefont {Sandev}},
  \bibinfo {author} {\bibfnamefont {L.}~\bibnamefont {Kocarev}},\ and\ \bibinfo
  {author} {\bibfnamefont {A.}~\bibnamefont {Pal}},\ }\href
  {https://doi.org/10.1103/PhysRevE.104.014121} {\bibfield  {journal} {\bibinfo
   {journal} {Phys.~Rev.~E}\ }\textbf {\bibinfo {volume} {104}},\ \bibinfo
  {pages} {014121} (\bibinfo {year} {2021})}\BibitemShut {NoStop}%
\bibitem [{\citenamefont {Donsker}(1951)}]{Donsker51}%
  \BibitemOpen
  \bibfield  {author} {\bibinfo {author} {\bibfnamefont {M.~D.}\ \bibnamefont
  {Donsker}},\ }\href@noop {} {\bibfield  {journal} {\bibinfo  {journal}
  {Mem.~Am.~Math.~Soc.}\ }\textbf {\bibinfo {volume} {6}},\ \bibinfo {pages}
  {1} (\bibinfo {year} {1951})}\BibitemShut {NoStop}%
\bibitem [{\citenamefont {Taqqu}(1975)}]{Taqqu75}%
  \BibitemOpen
  \bibfield  {author} {\bibinfo {author} {\bibfnamefont {M.~S.}\ \bibnamefont
  {Taqqu}},\ }\href {https://doi.org/10.1007/BF00532868} {\bibfield  {journal}
  {\bibinfo  {journal} {Z.~Wahrscheinlichkeitstheorie verw.~Gebiete}\ }\textbf
  {\bibinfo {volume} {31}},\ \bibinfo {pages} {287} (\bibinfo {year}
  {1975})}\BibitemShut {NoStop}%
\bibitem [{\citenamefont {Enriquez}(2004)}]{Enriquez04}%
  \BibitemOpen
  \bibfield  {author} {\bibinfo {author} {\bibfnamefont {N.}~\bibnamefont
  {Enriquez}},\ }\href {https://doi.org/10.1016/j.spa.2003.10.008} {\bibfield
  {journal} {\bibinfo  {journal} {Stoch.~Process.~Appl.}\ }\textbf
  {\bibinfo {volume} {109}},\ \bibinfo {pages} {203} (\bibinfo {year}
  {2004})}\BibitemShut {NoStop}%
\bibitem [{\citenamefont {Okamura}(2025)}]{Okamura25z}%
  \BibitemOpen
  \bibfield  {author} {\bibinfo {author} {\bibfnamefont {K.}~\bibnamefont
  {Okamura}},\ }\href {https://doi.org/10.5281/zenodo.17000139} {\bibinfo
  {title} {Data and figures for ``Fractional stochastic model of citation 
  dynamics with memory and volatility''}},\ \bibinfo
  {howpublished} {[Data set]} (\bibinfo {year} {2025}),\ \bibinfo {note}
  {{Zenodo}}\BibitemShut {NoStop}%
\bibitem [{\citenamefont {Garfield}(1975)}]{Garfield75}%
  \BibitemOpen
  \bibfield  {author} {\bibinfo {author} {\bibfnamefont {E.}~\bibnamefont
  {Garfield}},\ }\href
  {https://garfield.library.upenn.edu/essays/v2p396y1974-76.pdf} {\bibfield
  {journal} {\bibinfo  {journal} {Current Contents}\ }\textbf {\bibinfo
  {volume} {51/52}},\ \bibinfo {pages} {5} (\bibinfo {year}
  {1975})}\BibitemShut {NoStop}%
\end{thebibliography}
\end{document}